\documentclass[
 reprint,
 amsmath,
 pre,
 amssymb,
 nofootinbib,
 aps,
 longbibliography,
]{revtex4-1}
\usepackage[pdftex]{graphicx}
\usepackage{wasysym}
\usepackage{amsfonts}
\usepackage{ifsym}
\usepackage{pifont}
 \usepackage{footmisc}
\usepackage[export]{adjustbox}
\usepackage{dsfont}
\usepackage{lipsum}

\let\oldaddcontentsline\addcontentsline

\newcommand{\starttocentries}{\let\addcontentsline\oldaddcontentsline}
\makeatletter
\newlength{\apb@width}
\newcommand{\autoparbox}[2][c]{\settowidth{\apb@width}{#2}\parbox[#1]{\apb@width}{#2}}

\makeatother
\newcommand{\namedref}[2]{\hyperref[#2]{#1~\ref*{#2}}}

\newcommand{\Tr}{\operatorname{Tr}}

\newcommand{\Csphere}{{}^\bullet\kern-1.2pt C}
\newcommand{\Ctorus}{{}^\circ\kern-1.2pt C}

\newcommand{\nn}{\nonumber}

\newcommand{\COMMENT}[1]{}

\newcommand{\neqa}{\nonumber\end{eqnarray}}
\newcommand{\la}[1]{\label{#1}}

\newcommand{\<}{{\langle}}
\renewcommand{\>}{{\rangle}}

\newcommand{\re}{\relax{\rm I\kern-.18em R}}

\def\su2{{SU(2)}}

\def\[{\left[}
\def\]{\right]}

\def\({\left(}
\def\){\right)}
\def\[{\left[}
\def\]{\right]}

\def\<{\langle}
\def\>{\rangle}

\def\i2{\frac{i}{2}}

\def\2F1{\,_2{\rm F}_1}

\newcommand{\rrangle}{\rangle \hspace{-.15em} \rangle}
\newcommand{\llangle}{\langle\hspace{-.15em}\langle}

\usepackage[usenames,dvipsnames]{xcolor}
\usepackage{color}
\usepackage{graphicx}
\usepackage{dcolumn}
\usepackage{bm}
\usepackage{hyperref}

\usepackage{cancel}
\usepackage{ctable}
\usepackage{booktabs}
\newcolumntype{L}[1]{>{\raggedright\let\newline\\\arraybackslash\hspace{0pt}}m{#1}}
\newcolumntype{C}[1]{>{\centering\let\newline\\\arraybackslash\hspace{0pt}}m{#1}}
\newcolumntype{R}[1]{>{\raggedleft\let\newline\\\arraybackslash\hspace{0pt}}m{#1}}

\newcommand{\beq}{\begin{equation}}
\newcommand{\eeq}{\end{equation}}
\newcommand{\beqq}{\begin{equation*}}
\newcommand{\eeqq}{\end{equation*}}
\newcommand\beqa{\begin{eqnarray}}
\newcommand\eeqa{\end{eqnarray}}
\newcommand\beqaa{\begin{eqnarray*}}
\newcommand\eeqaa{\end{eqnarray*}}
\newcommand\bea{\begin{array}}
\newcommand\eea{\end{array}}

\begin{document}

%\preprint{APS/123-QED}

\title{Structure Constants in \texorpdfstring{$\mathcal{N} = 4$}{} SYM and Separation of Variables}

\author{Carlos Bercini}
\affiliation{Instituto de F\'isica Te\'orica, UNESP, ICTP South American Institute for Fundamental Research, Rua Dr Bento Teobaldo Ferraz 271, 01140-070, S\~ao Paulo, Brazil}
\author{Alexandre Homrich}
\affiliation{Department of Physics and Astronomy, University of Waterloo, Waterloo, Ontario, N2L 3G1, Canada \\
Perimeter Institute for Theoretical Physics, 31 Caroline St N Waterloo, Ontario N2L 2Y5, Canada}
\author{Pedro Vieira}
\affiliation{Perimeter Institute for Theoretical Physics, 31 Caroline St N Waterloo, Ontario N2L 2Y5, Canada
\\
Instituto de F\'isica Te\'orica, UNESP, ICTP South American Institute for Fundamental Research, Rua Dr Bento Teobaldo Ferraz 271, 01140-070, S\~ao Paulo, Brazil}

             
\begin{abstract}
We propose a new framework for computing three-point functions in planar $\mathcal{N}=4$ super Yang-Mills where these correlators take the form of multiple integrals of Separation of Variables type. We test this formalism at weak coupling at leading and next-to-leading orders in a non-compact SL(2) sector of the theory and all the way to next-to-next-to-leading orders for a compact SU(2) sector. We find evidence that wrapping effects can also be  incorporated. 
\end{abstract}

\pacs{Valid PACS appear here}

\maketitle

\section{Introduction}
Solving planar $\mathcal{N}=4$ SYM in a satisfactory way would mean efficiently computing both the spectrum as well as higher point correlation functions -- starting with three points -- at any value of the 't Hooft coupling.

A formalism for computing three point functions by means of integrability exists. It is the so called \textit{hexagon} approach \cite{hexagons}. It is conjectured to hold for any coupling indeed but it is not easy to use, at least not by the remarkable standards of the spectrum quantum spectral curve approach \cite{kolyaQSC}. With \textit{hexagons} one needs to go over infinitely many sums and integrals to produce such correlators. At weak coupling perturbation theory these sums and integrals truncate \cite{wrappingthreeBGKV, wrappingthreeES, wrappingshota} and we can use hexagons to produce a wealth of data to test any putative new framework. We will do it all the time below.

Here we suggest a new approach for correlation functions in $\mathcal{N}=4$ SYM based on the Baxter Q-functions. The final representations are of so-called separation of variables (SoV) type where these Baxter functions are integrated against simple universal measures to produce the structure constants.

%Recall that in the spectrum problem 
Q-functions play the central role in the quantum spectral curve, the top of the line tech for computing the dimension of any single trace operator in this conformal gauge theory so it is only natural to look for a similar central role for these objects in the context of other physical observables such as the OPE structure constants.  

In the most conventional integrable spin chains, Q-functions are polynomials whose roots are the so called Bethe roots $v_k$. In SYM these polynomials are present at leading order at weak coupling but at higher coupling they get dressed by quantum non-polynomial factors. This is expected; as we crank up the coupling we are no longer dealing with a spin chain or with a classical string but something in between and so these Baxter polynomials get naturally deformed. For the so-called SL(2) sector of the theory which includes all operators of the schematic form $\Tr(D^JZ^L)+\texttt{permutations}$ we have 
%\beqa
%\label{baxterDef} &&\mathcal{Q}({\color{blue}u}) \equiv \prod\limits_{k=1}^{J}\frac{{\color{blue}u}-v_k}{\sqrt{x^{+}(v_k)x^{-}(v_k)}} \times \\ &&\exp\Bigg( \frac{1}{2}g^2 Q_1^+ \big(H_1\left(-\tfrac{1}{2}+i \textcolor{blue}{u}\big)+H_1\big(-\tfrac{1}{2}-i \textcolor{blue}{u}\big)\right)  +\nonumber \\
%&& -\frac{i}{2} g^2 Q_1^- \left(H_1\big(-\tfrac{1}{2}+i \textcolor{blue}{u}\big)-H_1\big(-\tfrac{1}{2}-i \textcolor{blue}{u}\big)\right) + O(g^4)\Bigg) \nn 
%\eeqa
\beqa
\label{baxterDef} \mathcal{Q}({\color{blue}u}) \equiv\! \(\prod\limits_{k=1}^{J}\frac{{\color{blue}u}-v_k}{\sqrt{x_k^{+}x_k^{-}}} \)\!
%\times \\ &&\exp\Bigg( 
e^{\frac{g^2}{2} Q_1^+ H_1^+({\color{blue} u}) +\frac{g^2}{2}  Q_1^- H_1^-({\color{blue} u}) + O(g^4)}
%\Bigg) \nn 
\eeqa
where the charges $Q_1^\pm$ are simple functions of the Zhukowsky variables $x_k^\pm=x^\pm(v_k)$ and $H_k^\pm$ are harmonic functions, see appendix \ref{rootscharges}. 
%At higher loops, the dressing of the Baxter polynomials will probably depend on the geometry of the correlator. 

\begin{figure}[t]
    \centering
    \includegraphics[width=0.47\textwidth]{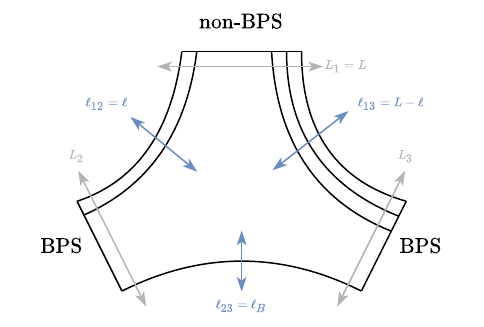}
   % \vspace{-1cm}
    \caption{Three point functions of operators with sizes $L_1$, $L_2$ and $L_3$. At tree level there are $\ell_{ij}=(L_i+L_j-L_k)/2$ propagators between operators $i$ and $j$; these integers $\ell_{ij}$ are called the bridge lengths. We will usually use $L\equiv L_1$ for the length of the non-BPS operator and $\ell \equiv \ell_{12}$ for its  so-called left adjacent bridge (the right adjacent bridge will have length $L-\ell$); the bottom bridge we will often denote as $\ell_{B}\equiv \ell_{23}$. 
    %The unprotected operator is connected with the other two BPS operators via two bridges, the left one is $\ell = (L+L_2-L_3)/2$ and the right one is $L-\ell$.}
    }
    \label{lBridge}
\end{figure}

An operator is given by a Q-function (\ref{baxterDef}). What we are after is thus a functional eating these functions and spitting out a number, the structure constant.

For the most part we will consider a single non-BPS operator and two BPS operators. The geometry of the three point function depends on the size of all these operators or -- equivalently -- on the so-called bridges that connect them as reviewed in figure \ref{lBridge}.  
%\textbf{[Re-write: Explain what bridges are adding a figure]} We could consider, for example, a three-point function involving one non-BPS SL(2) operator of finite twist $L$ and two BPS operators of sizes $L_2$ and $L_3$. As illustrated in figure \ref{lBridge} there is an important integer $\ell=(L+L_2-L_3)/2$ parametrizing this three-point function. It is the so called \textit{left bridge}. There is also a \textit{right bridge} of length $L-\ell$ as illustrated in that figure. If $L_2$ and $L_3$ is large with their difference finite (so that $\ell$ is finite) the structure function is a function of the length $L$ and of the baxter function $\mathcal{Q}$ parametrizing the non-BPS operator but also of $\ell$ parametrizing the three-point function geometry. 

The proposal is that (the square of) the structure constant is given by a ratio of SoV like scalar products
%\begin{equation}
%C_{\bullet\circ\circ}^2 = \binom{2\mathbb{J}+L-1}{\mathbb{J}}^{-1}\frac{\langle \mathcal{Q},\mathbf{1}\rangle_{\ell}^2}{\langle\mathcal{Q},\mathcal{Q}\rangle_{L}}
%\label{AbelianOneSpin}
%\end{equation}
\begin{equation}
C_{\bullet\circ\circ}^2 = \frac{(\mathbb{J}!)^2}{(2\mathbb{J})!}\frac{\langle \mathcal{Q},\mathbf{1}\rangle_{\ell}^2}{\langle\mathcal{Q},\mathcal{Q}\rangle_{L}}
\label{AbelianOneSpin}
\end{equation}
Here $\mathbb{J} = J+ g^2 Q_1^+$ and the scalar product
\begin{align}
    \langle \mathcal{Q}_1,\mathcal{Q}_2 \rangle_\ell & \equiv \binom{\mathbb{J}_1+\mathbb{J}_2+\ell-1}{\ell-1}\!\int\! d\mu_{\ell-1}\prod_{i=1}^{\ell-1} \mathcal{Q}_1(u_i)\mathcal{Q}_2(u_i) \la{scalarProduct}
\end{align}
with a nice factorized measure
\begin{equation}
d\mu_{\ell} = \prod_{i=1}^\ell du_i \, \mu_1(u_i)\prod_{i=1}^{\ell-1}\prod_{j=i+1}^{\ell}\mu_2(u_i,u_j)
\label{multimeasure}
\end{equation} 
which is constructed out of the building blocks (using, $s_u,c_u,t_u$ for $\sinh(\pi u),\cosh(\pi u),\tanh(\pi u)$)
\beqa
&&\!\!\!\!\!\!\!\!\!\!\!\!\!  \mu_1(u)\! =\! \frac{\pi}{2 c_u^2}\left(1+ g^2 \pi^2\!\left(3 t_u^2-1\right)\!+ ...\right) \label{onePMeasure} 
\\
&&\!\!\!\!\!\!\!\!\!\!\!\!\!  \mu_2(u,v)\! =\! \frac{\pi(u-v)s_{u-v}}{c_u c_v} \label{twoPMeasure} 
%\times \\&& 
\!\left(1\!+\!g^2 \pi^2\!\left((t_u+t_v)^2\!-\!\tfrac{4}{3}\right)\! +\! ...\right) %\nonumber
\eeqa
%\beqa
%&&\!\!\!\!\!\!\!\!\!\!  \mu_1(u)\! =\! \frac{\pi/2}{\cosh^2(\pi u)}\left(1+\pi^2 g^2\!\left(3\tanh^2(\pi u)-1\right)\!+ \dots\right) \label{onePMeasure} 
%\\
%&&\!\!\!\!\!\!\!\!\!\!  \mu_2(u,v) = \frac{\pi(u-v)\sinh(\pi(u-v))}{\cosh(\pi u)\cosh(\pi v)} \label{twoPMeasure} \times \\&& \left(1+g^2\frac{\pi^2}{3}\left(3(\tanh(\pi u)+\tanh(\pi v))^2-4\right) + \dots\right) \nonumber
%\eeqa
valid to leading order (LO) and next-to-leading order (NLO); the dots in the above expressions are NNLO corrections which start at $O(g^4)$. The structure of the result might be corrected at subleading orders as explained below. The contours in (\ref{AbelianOneSpin}) are the real axis for all~$u_i$.

%\textcolor{red}{We must comment on the invariance of left vs right. In this notation is simply $\langle \mathcal{Q},1\rangle_{\ell} = \langle \mathcal{Q},1\rangle_{L-\ell}$, and also comment in how non-trivial this equality is. Should we do it here? P: Isn't it in page 3? Please read the full draft... Or did you mean something else/saw it but wanted it here etc? C: I thought here was better because we could easily ref. the figure }

We can \textit{prove (\ref{AbelianOneSpin}) by exhaustion} by comparing it with hexagon produced data for numerous $L$'s and $\ell$'s and for  different operators corresponding to different $\mathcal{Q}$-functions. We did it; (\ref{AbelianOneSpin}) is correct. We can also establish it more honestly as discussed in the next section. 

Representations like (\ref{AbelianOneSpin}) are the main results of this letter. In section \ref{SU2comments} we present an SU(2) counterpart of this representation in (\ref{su2structure}); we managed to fully test it to LO, NLO \textit{and} NNLO. We compare these two rank one sectors in section \ref{comparison}. In the discussion section \ref{discussion} we discuss further generalizations such as multiple spinning operators and speculate on all loop structures we expect to find. Many appendices complement the main text. 

\section{SL(2)} \label{SL2comments}
At leading order, that is at tree level, correlation functions are given by Wick contractions. Each operator can be through of as a spin chain and these Wick contractions are thus given by spin chain scalar products \cite{tailoring0,tailoring,Wang}. For the SL(2) sector such scalar products can be cast as SoV integrals \cite{Derkachov} -- see also \cite{clustering} in the $\mathcal{N}=4$ context. Once we properly normalize all these scalar products as in \cite{Wang} to extract the structure constant we precisely end up with~(\ref{AbelianOneSpin}) for~$g=0$. 

What we find remarkable is that this classical $g=0$ expression seems to have a nice quantum lift as anticipated in the introduction. 

Let us first discuss the simplest possible case where a twist two operator ($L=2$) splits evenly into two BPS operators ($\ell=1$) so that the proposal (\ref{AbelianOneSpin}) simply reads
\beq
C_{\bullet\circ\circ}^2 = \frac{\mathbb{J}!^2}{(2\mathbb{J}+1)!} \( 
\int du\, \mu_1(u) \mathcal{Q}(u)^2 \)^{-1} \,. \la{twist2C123}
\eeq
We derived the single-particle measure $\mu_1$ as well as the quantum deformed Baxter functions $\mathcal{Q}$ in two ways. Both are based on looking for a measure such that, for two different twist-two states (i.e. with different spins $J$ and $J'$), we have an orthogonality relation 
\beq
\int \mu_1(u) \mathcal{Q}_J(u) \mathcal{Q}_{J'}(u)  \propto \delta_{JJ'}  \,, \la{orthoSL2}
\eeq
a powerful relation which has been extensively exploited by \cite{kolyasl3, kolyasu2, kolyadeterminant} in numerous SoV studies, most of which for rational spin chains or for the fishnet reduction \cite{fishnet} of SYM, see most notably \cite{kolyafishnet}.

The first uses the fact that the Bethe roots $v_k$ of twist two operators are given by the zeros of Hahn polynomials, a fact that persists at NLO. We have that $\prod_{k} ({\color{blue} u}-v_k)$ is proportional to the Hahn polynomial \cite{LOHahn,NLOHahn} $p_J(u|a,b,b,a)$ where $a-b=2 i \sqrt{2} g$ and $a+b=1+4g^2 H_1(J)$.
Hahn polynomials are orthogonal with a simple measure as reviewed in appendix \ref{HahnDetails} which allows us to derive (\ref{onePMeasure}) up to some simple tuning due to the mild $J$ dependence in the polynomial parameters $a,b$, see appendix for details. It is the $J$ dependence that renders the derivation non-trivial and which is responsible for the needed modification of the Baxter functions in~(\ref{baxterDef}). 

The second derivation follows \cite{kolyasl3} closely (the novelty being the extension to $g^2$ corrections) and makes use of the Baxter equation \cite{bassobelitsky}
\beq
\mathcal{B} \circ \mathbb{Q}=T(u) \mathbb{Q}(u) \label{baxterT}
\eeq
where $\mathbb{Q}(u)$ are the Baxter polynomials (i.e. just the parentheses in (\ref{baxterDef})) and the Baxter operator 
\beq
\mathcal{B} \equiv (x^+)^L \Big(1- \frac{g^2}{x^-(u)}(Q_1^+ - i Q_1^-)\Big) e^{ i \partial_{u}}  + c.c.
\eeq
Note that at $g=0$ we have $x^+=u+i/2$ and most importantly, $\mathcal{B}$ becomes a simple linear operator but as we turn on $g$ corrections this is no longer true since the charges $Q_1^\pm$ depend on $\mathbb{Q}$. Consider first $g=0$ and $L=2$ so that the transfer matrix is $T(u)=2u^2+c_J$.  Multiplying (\ref{baxterT}) by the sought after measure $\mu_1$ and by another Baxter polynomial with a different spin, subtracting that to the same thing with the spin swapped and integrating yields 
\beq
(c_J-c_{J'} )\! \int \! \mu_1  \mathbb{Q}_J \mathbb{Q}_{J'} \!=\!\!\int \!\mu_1\big( \mathbb{Q}_{J'} (\mathcal{B} \circ \mathbb{Q}_J)- (\mathcal{B} \circ \mathbb{Q}_{J'}) \mathbb{Q}_{J}  \big). \la{baxterDerivation}
\eeq
If we manage to make $\mathcal{B}$ self-adjoint we will thus have the required orthogonality. An $i$-periodic $\mu_1$ would do the job since under shifts of contour the two terms of Baxter would swap and cancel in the right hand side. In detail, 
$\mathbb{Q}_{J'}(u)(u\pm \tfrac{i}{2})^2 \mathbb{Q}_J(u\pm i) \mu_1(u)$ becomes $\mathbb{Q}_{J'}(u\mp i)(u\mp \tfrac{i}{2})^2 \mathbb{Q}_J(u) \mu_1(u\mp i)$  under a shift of contour by $\mp i$ leading to an interchange (and thus cancellation) of the two terms in  (\ref{baxterDerivation}) once we use that the measure is periodic.
%becoming nothing but one of the terms in the second term in the right hand side of (\ref{baxterDerivation}) once we use that the measure is periodic. 
To make this manipulation kosher we need to make sure no singularities are picked when deforming the contour and to make the measure acceptable we need to make sure it decays fast enough at infinity so that it can be integrated against polynomials of arbitrary degree. Both this problems are solved at once with 
\beq
\mu_1(u)= \frac{\pi/2}{\cosh^2(\pi u)} + O(g^2) \,. \la{cosh2}
\eeq
The function decays faster than any polynomial and the double poles at $\pm i/2$ precisely cancel the double zeroes in the potential terms $(u \mp i/2)^L$ when $L=2$ so that they lead to no extra contribution when deforming the contours. A periodic function without these double poles would not decay fast enough and a function with more than double poles would lead to extra contributions when deforming the contours;~(\ref{cosh2}) is the sweet spot. 

Turning on $g$ corrections is not complicated. The redefinition (\ref{baxterDef}) brings the Baxter operator to a linear operator again but introduces some further poles at $\pm i/2$ in the (no longer polynomial) Baxter $\mathcal{Q}$ functions so that the measure now needs some extra poles to cancel the contribution of these when deforming the contour. This explains (\ref{baxterDef}) as well as (\ref{onePMeasure}); more details in appendix~\ref{BaxterSL2Ap}. 

Having derived the measure $\mu_1$ and the Baxter polynomial dressing it remains to fix the overall normalization of the structure constant to be sure everything is in order. Evaluating the SoV integral using the loop corrected Bethe roots for various spins immediately leads to the cute result
\beq
 \( 
\int du\, \mu_1(u) \mathcal{Q}(u)^2 \)^{-1}\!\! = \frac{1+g^2 \big(4 H_2(J)+\frac{8H_1(J)}{2J+1}\big)+\dots }{2J+1}
\eeq
which is indeed almost the correct loop level structure constant computed in \cite{DO}. The prefactor in (\ref{twist2C123}) with $J$ deformed into $\mathbb{J}$ in a \textit{reciprocity} reminiscent fashion~\cite{reciprocitypapers} neatly combines with this expression to give the full NLO structure constants for twist-two operators. 

At this point it is straightforward to guess the general structure constant for any SL(2) operators of any twists by simply taking the tree level $g=0$  result and deforming the new ingredient for higher twists -- the two particle measure $\mu_2$ -- by a bunch of hyperbolic tangents following what was derived for twist two. The coefficients of these hyperbolic tangents are then fixed by requiring orthogonality between any two different Baxter solutions. 
%We also needed to guess 
The last line deformation of the Baxter functions in (\ref{baxterDef}) -- which was invisible for twist two operators for which odd charges vanish -- is also fixed by imposing orthogonality. In the end, we just need to check  that with a minimal reciprocity friendly prefactor as in~(\ref{AbelianOneSpin}) we precisely agree with perturbative data produced by hexagons. We do.

Even without matching with data, there is a nice self-consistency check of the full construction including the deformed pre-factor: The structure should be invariant under swapping left and right bridges $\ell \leftrightarrow L-\ell$; we checked that this is indeed realized by our expressions once the prefactors are deformed as in (\ref{AbelianOneSpin}). 
%Another comment is that the corrections to the measure and Baxter functions organize themselves nicely transcendentality wise.

We made some progress at higher loops, in particular at NNLO (two loops) and for the smallest possible sizes and bridge lengths. For twist $L=2$ operators for example we found a Baxter function dressing as well as an orthogonal one-particle measure realizing (\ref{orthoSL2}) as 
\begin{align}
    \mathcal{Q}({\color{blue}u}) & \equiv \prod\limits_{k=1}^{J}\frac{{\color{blue}u}-v_k}{\sqrt{x^{+}_kx_k^{-}}}\times e^{\frac{1}{2}g^2 Q_1^{+}H_1^{+}+\frac{1}{8}g^4 Q_1^{+}Q_2^{+}H_1^{+}-\frac{1}{2}g^4 Q_1^{+}H_3^{+}}\nonumber
\end{align}
and
%\begin{align}
%\mu_1(u) &= \frac{\pi/2}{\cosh^2(\pi u)}\Big(1+\pi^2 g^2 \big(3\tanh^2(\pi u)-1 \big)+ \nonumber \\
%&\!\!\!\!\!\!+\pi^4g^4\big(\tfrac{5}{6}-7\tanh^2(\pi u)+\tfrac{11}{2}\tanh^4(\pi u)\big)+\nonumber\\
%&\!\!\!\!\!\!-\frac{g^4}{8}H_1^{+} \big(Q_1^{+}(\mathbf{v}_1)Q_2^{+}(\mathbf{v}_2)+Q_1^{+}(\mathbf{v}_2)Q_2^{+}(\mathbf{v}_1)\big) \Big)  \label{onePMeasureNNLO} \,.
%\end{align}
\begin{align}
\!\!\mu_1(u) &= \frac{\pi}{2c_u^2}\Big(1+\pi^2 g^2 \big(3t_u^2-1 \big)+ \pi^4g^4\big(\tfrac{5}{6}-7t_u^2+\tfrac{11}{2}t_u^4\big)\nonumber\\
&\!\!\!\!\!\!-\frac{g^4}{8}H_1^{+} \big(Q_1^{+}(\mathbf{v}_1)Q_2^{+}(\mathbf{v}_2)+Q_1^{+}(\mathbf{v}_2)Q_2^{+}(\mathbf{v}_1)\big) \Big)  \label{onePMeasureNNLO} \,,
\end{align}
where we 
%introduce the notation $H_n^{+}=H_n\left(-1/2+iu\right)+H_n\left(-1/2-i u\right)$ and 
use the fact that $Q^{-}_k =0$ for twist-2 operators.

%which we find quite interesting. 
The last line is exotic as it depends now on the charges of the two operators in (\ref{orthoSL2}) with Bethe roots $\mathbf{v}_1$ and $\mathbf{v}_2$ respectively. Note, however, that when considering the pairings $\langle\mathcal{Q}, \textbf{1} \rangle$ and $\langle\mathcal{Q}, \mathcal{Q} \rangle$ a single Baxter function shows up and thus this mixing term can be absorbed as new factor dressing $\mathcal{Q}$. One would then have different dressings in each pairing, a phenomenon we observe in the next section as well (in the SU(2) sector). 

\begin{figure}
    \centering
    \includegraphics[width=0.55\textwidth]{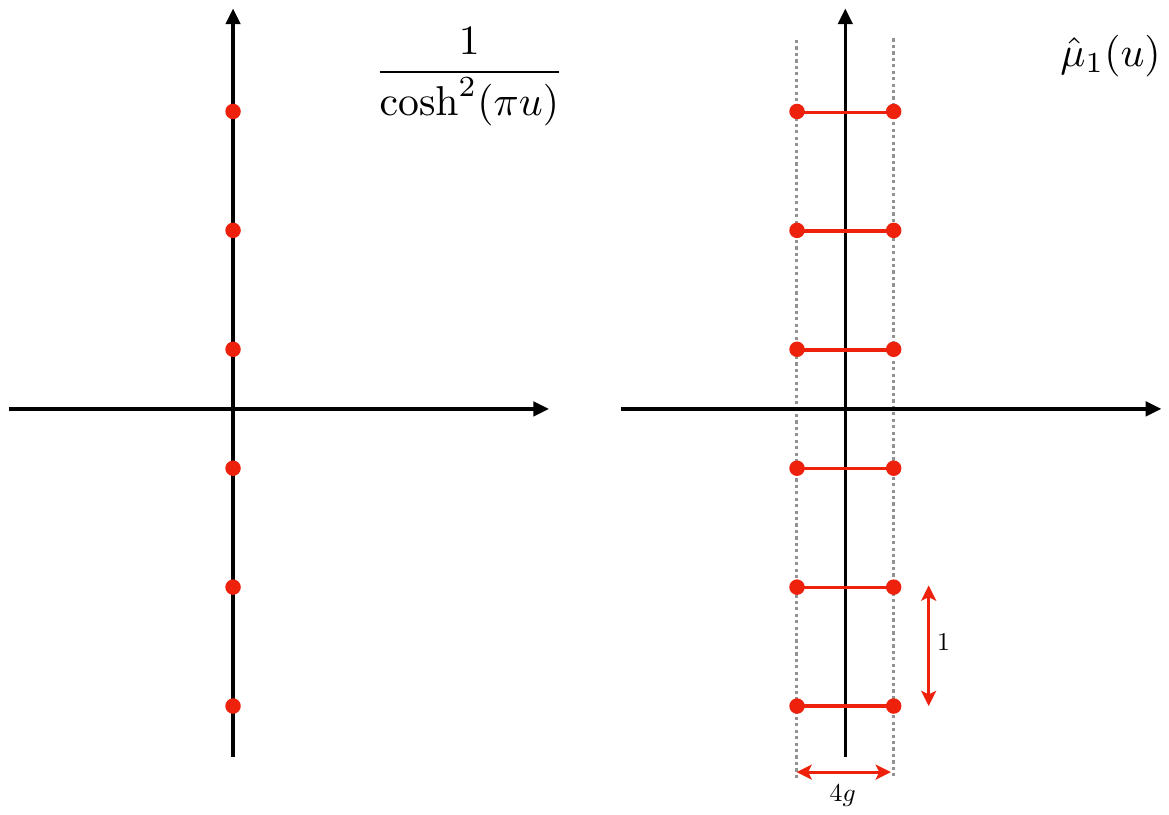}
    \vspace{-3cm}
    \caption{\textit{Zhukowskization} of trigonometric functions opens up an infinite tower of poles into an infinite ladder of cuts. }
    \label{polesCuts}
\end{figure}

The first line in (\ref{onePMeasureNNLO}) is also not any random combination of trigonometric functions. Take the tree level measure  $1/\cosh^2(\pi u)$ -- which is periodic with period $i$ and has poles at all the imaginary half integers -- and promote it to a periodic function where all these poles are opened up into small cuts following~\cite{kolyafishnet}, see figure \ref{polesCuts}, 
\beq
\hat{\mu}_1  \equiv \oint \frac{dv}{2\pi i} \frac{\pi/2}{\cosh^2({\pi (u-v)})} \frac{1}{x(v)}  \,. \la{muhat}
\eeq
The integration contour encircles the Zhukowsky cut~$v \in [-2g,2g]$. Evaluating this integral in perturbation theory precisely reproduces the first line in (\ref{onePMeasureNNLO}) up to an overall normalization constant! It is tempting to conjecture that the finite coupling expression (\ref{muhat}) might play an important role in an all loop SoV formulation. We make further comments on NNLO structure constants in the discussion.

%{\color{red}P: Should we remove the next paragraph?}
%We made some progress in correcting the structure constant itself at NNLO but we did not manage to fix all the normalization factors therein and write a clean final expression valid for all sizes and bridge lengths. We discuss some of this in the discussions. 

%[refer to appendices bringing perhaps one formula for Gaudin and one for calA at l=1?]
%At higher loops we expect the dressing in (\ref{baxterDef}), the quantum corrections to the measures (\ref{onePMeasure}) and (\ref{twoPMeasure}) and the full structure constant to follow a similar expansion with a clean transcendentality pattern. We tried guessing it and failed, see appendix \ref{orthoFailingL2} for an example. Hopefully we are simply overlooking some simple new ingredient which will make it all work. 
%It is also worth checking 
Other sectors such as the SU(2) sector might also hint at other important structures in a putative all loop formulation. This is what we turn to now. 

\section{SU(2)} \label{SU2comments}
Up to NNLO the SU(2) magnon S-matrix does not receive corrections. Quantum effects do correct the propagation of scalar particles on top of the BMN vacuum. One could imagine mimicking this dressed propagation through appropriate deformations of the background vacuum by the insertions of inhomogeneities $\theta_i$. This was made concrete in \cite{morphism} where the authors determined NLO (one loop) structure constants in the SU(2) sector through the construction of a differential ``$\theta$-morphism" operator acting on the ($\theta$-deformed) spin-chain scalar products defining the LO structure constants \cite{tailoring} and outputting the quantum three point functions.

Here we make two observations. First we note that there exists a $\theta$-morphism that promotes XXX$_{1/2}$ inhomogenous spin chain scalar products,
\begin{align}
\mathcal{A}_{\theta} &= \sum_{\textbf{u} = \alpha\cup\bar{\alpha}} (-1)^{|\bar{\alpha}|}\prod_{n =1}^{\ell} \prod_{i \in \bar{\alpha}}  \frac{u_i - \theta_n +i/2 }{u_i - \theta_n -i/2 } \prod_{j \in \alpha} \frac{u_j - u_i + i}{u_j-u_i},\nonumber\\
\mathcal{B}_{\theta} &= \det\left[\partial_{u_i}\log\left(\prod_{n =1}^{L}  \frac{u_j - \theta_n +i/2 }{u_j - \theta_n -i/2 }\prod_{k\neq j} \frac{u_j - u_k - i}{u_j-u_k +i} \right)\right] \nonumber\\& \hspace{0.5cm}\times \prod_{i < j} \frac{(u_i - u_j)^2}{1+ (u_i - u_j)^2}, \label{gaudinwiththeta}
\end{align}
into $\mathcal{N}=4$ SU(2) structure constants all the way to NNLO:
\begin{align}
C^2_{\bullet \circ \circ} = \left|\frac{\left(\mathcal{M} \circ \mathcal{A}_\theta\right)^2}
{\Lambda_\mathcal{B} \, \mathcal{M} \circ \mathcal{B}_\theta}\right|\Bigg|_{\theta = 0} + O(g^6), \label{morphismaction}
\end{align}
where
\begin{align}
\mathcal{M} =& \exp  \Bigg[ \sum_{i=1}^L  \left(g^2  (\partial_{i, i+1})^2  - \frac{1}{4} g^4  (\partial_{i,i+1})^2(\partial_{i+1,i+2})^2\right) \nonumber\\
& -i g^2  Q^+_1 (\partial_{1}  - \partial_{L}) + g^4 \delta \mathcal{M}_\text{NNLO-b}\Bigg], \label{morphismeq}
\end{align}
with $\partial_i = \partial_{\theta_i}$ and $\partial_{i,i+1} = \partial_i - \partial_{i+1}$, see appendix \ref{morphismappendix} for $\delta\mathcal{M}_\text{NNLO-b}$, and where $\Lambda_\mathcal{B}$ is a simple norm factor, (\ref{LambdaB}). To NLO this is just the construction of \cite{morphism}. We fix the NNLO part through comparison with the hexagons prediction, see appendix \ref{morphismappendix} for details. 

The second observation is that given the known integral representation for the scalar products in the inhomogeneous XXX$_{1/2}$ spin-chain, derived from Sklyanin's~\cite{shotaXXX}, Baxter's \cite{kolyasu2} and hexagon methods \cite{shotamirror,clustering}, one can straightforwardly derive the functional SOV representation for the NNLO structure constants simply by acting with the $\theta$-morphism operator on this classical measure. 

The result is once again (the square of) a $\ell-1$ dimensional integral over an $L-1$ dimensional integral,
%\begin{align}
%C^2_{\bullet \circ \circ} = \frac{\binom{l}{J}^2\lambda_\ell(\mathcal{Q}) }{\binom{L}{J}\binom{L-J}{J}} \frac{\llangle \mathcal{Q},\mathbf{1}\rrangle^2_{\ell,L}}{\llangle \mathcal{Q},\mathcal{Q}\rrangle_{L,L}} + O(g^6), \label{su2structure}
%\end{align}
\begin{align}
C^2_{\bullet \circ \circ} =\Lambda_\ell(\mathcal{Q}) \times \frac{(J!)^2}{(2J)!} \times \frac{\llangle \mathcal{Q},\mathbf{1}\rrangle^2_{\ell,L}}{\llangle \mathcal{Q},\mathcal{Q}\rrangle_{L,L}} + O(g^6), \label{su2structure}
\end{align}
with $\llangle f, g \rrangle_{\ell,L} \equiv \langle f, g \rangle_{\ell, L}/ \langle \textbf{1}, \textbf{1} \rangle_{\ell, L}$ and
\beq
 \langle \mathcal{Q}_1, \mathcal{Q}_2 \rangle_{\ell, L} \equiv \binom{\ell}{J_1+J_2}\oint\limits_\gamma d\mu_{\ell, L}  \prod_{i=1}^{\ell-1} \mathcal{Q}_1(u_i) \mathcal{Q}_2(u_i). 
\label{pairingsu2}\eeq
%\beq
% \langle f, g \rangle_{\ell, L} \equiv \oint_\gamma d\mu_{\ell, L}  \prod_{i=1}^{\ell-1} f(u_i) g(u_i). 
%\eeq
The contour of integration is a circle wrapping the singularities of the measure (i.e. both Zhukowsky cuts), which once again is factorized:
\beq
d\mu_{\ell,L}=  \prod_{i=1}^{\ell-1} du_i\,  \mu^1_{\ell,L}(u_i) \prod_{j\neq i}^{\ell-1}\mu_{\ell,L}^2(u_i,u_j), \label{periodicmeasuresu2}
\eeq
with
\begin{align}
&\mu^1_{\ell,L}(u) \label{measure1su2}= \frac{\sinh(2\pi u)}{(x^+_u x^-_u)^2} e^{\delta_{\ell \neq L} \texttt{A}_{\ell,L}(u)}
\\
&\mu_{\ell,L}^2(u,v) \label{measure2su2}=\frac{\sinh(2\pi (u-v))(u-v)}{2 x^+_u x^-_u x^+_v x^-_v} e^{\delta_{\ell \neq L} \texttt{A}_{\ell,L}(u,v)}\\
&\mathcal{Q}({u}) \equiv \prod\limits_{k=1}^{J}\frac{{u}-v_k}{\sqrt{x^{+}_k x^{-}_k}}e^{\delta_{\ell \neq L} \texttt{B}(u)} 
\end{align}
The scalar product in the denominator of (\ref{su2structure}) enters with $\ell=L$, and therefore the exponential in all these expressions should be dropped in that case. For the numerator we need these extra quantum dressing factors. They read
\beqa
&&\!\!\!\!\!\!\!\texttt{A}_{\ell,L}(u)=g^2 (q^-_1)^2  + g^4 \left( \tfrac{1}{2}(q^+_2)^2 + 4 \alpha (q^-_1)^2 - 6\pi^2  q^+_2  {\color{red}{\delta_{\ell=2}}}\right) \nn\,,\\
&&\!\!\!\!\!\!\!\texttt{A}_{\ell,L}(u,v)=g^2 q^-_1 \tilde{q}^-_1  + g^4 \left( \tfrac{1}{2}q^+_2 \tilde{q}^+_2 + 2\alpha  q^+_2 + 4 \alpha q^-_1\tilde{q}^-_1 \right) \,,\nn\\
&&\!\!\!\!\!\!\!\texttt{B}(u)=g^2 q^-_1 Q^-_1 \label{dressing} \\
&&\!\!\!\!\!\!\!+ g^4 \Big( \tfrac{1}{2}q^+_2 Q^+_2 + \alpha  Q^+_2 + 4\alpha q^-_1 Q^-_1  \nonumber - \pi^2q^-_1 Q^-_1 {\color{red}{\delta_{\ell=2}}}  +\nn\\
&&\!\!\!\!\!\!\!+\big(\big(\tfrac{1}{8} ({Q_1^+})^2\! -\!\tfrac{1}{4}  Q_2^+\! - \! Q_1^+\! +\!\tfrac{3}{8}  ({Q_1^-})^2\big) q_2^+\! -\!\tfrac{1}{2} q_3^- Q_1^-\big){\color{blue}{\delta_{\ell,L-1}}}\Big) \,, \nn
\eeqa
%\begin{align}
%&\mu^1_{\ell,L}(u) \label{measure1su2}= \frac{\sinh(2\pi u)}{(x^+_u x^-_u)^2}\times\\&\exp\left[\delta_{\ell \neq L}\left(g^2 (q^-_1)^2  + g^4 \left( \tfrac{1}{2}(q^+_2)^2 + 4 \alpha (q^-_1)^2 - 6\pi^2  q^+_2 {\color{red}{\delta_{\ell=2}}}\right)  \right)\right] \nonumber
%\\&\mu_{\ell,L}^2(u,v) \label{measure2su2}=\frac{\sinh(2\pi (u-v))(u-v)}{2 x^+_u x^-_u x^+_v x^-_v}\times\\ \nonumber& \exp\left[\delta_{\ell \neq L}\left(g^2 q^-_1 \tilde{q}^-_1  + g^4 \left( \tfrac{1}{2}q^+_2 \tilde{q}^+_2 + 2\alpha  q^+_2 + 4 \alpha q^-_1\tilde{q}^-_1 \right) \right) \right] ,
%\end{align}
where the charges $q^{\pm}_k =q^{\pm}_k(u)$ and $\tilde{q}^{\pm}_k =q^{\pm}_k(v)$ are defined in appendix \ref{notation} and $\alpha = \tfrac{2}{3} \pi^2 -1$. 
%The Q-functions are dressed \cite{dressingsu2gaudin} as
%\begin{align}
%\label{dressingsu2} &\mathcal{Q}({u}) \equiv \prod\limits_{k=1}^{J}\frac{{u}-v_k}{\sqrt{x^{+}_k x^{-}_k}}\exp\Bigg[\delta_{\ell \neq L}\Bigg(g^2 q^-_1 Q^-_1  + g^4 \Big( \tfrac{1}{2}q^+_2 Q^+_2 + \\& \alpha ( Q^+_2 + 4 q^-_1 Q^-_1)  \nonumber - \pi^2q^-_1 Q^-_1 \delta_{\ell=2}  - \tfrac{1}{2} \left( q_2^+ Q_2^+ + q_3^- Q_1^-\right)\delta_{\ell,L-1}\Big) \Bigg)\Bigg] \nonumber
%\end{align}
%\beqa
%&&\!\!\!\!\!\!\!\label{dressingsu2} \mathcal{Q}({u}) \equiv \prod\limits_{k=1}^{J}\frac{{u}-v_k}{\sqrt{x^{+}_k x^{-}_k}}\times\exp\Bigg[\delta_{\ell \neq L}\Bigg(g^2 q^-_1 Q^-_1 + \\
% &&\!\!\!\!\!\!\!+ g^4 \Big( \tfrac{1}{2}q^+_2 Q^+_2 + \alpha  Q^+_2 + 4\alpha q^-_1 Q^-_1  \nonumber - \pi^2q^-_1 Q^-_1 {\color{red}{\delta_{\ell=2}}}  +\\&&\!\!\!\!\!\!\!+\big(\big(\tfrac{1}{8} {Q_1^+}^2 -\tfrac{1}{4}  Q_2^+ -  Q_1^+ +\tfrac{3}{8}  {Q_1^-}^2\big) q_2^+ -\tfrac{1}{2} q_3^- Q_1^-\big){\color{blue}{\delta_{\ell,L-1}}}\Big) \Bigg)\Bigg] \nonumber
%\eeqa
Finally $\Lambda_\ell(Q)$ is a simple function of the higher conserved charges also given in appendix \ref{notation}. 
%Note that the scalar product in the denominator of (\ref{su2structure}) enters with index $\ell=L$, and therefore the exponential in all these expressions should be dropped in that case.

%\beq
%\Lambda_\ell(\mathcal{Q})= \frac{\exp \left[2g^4\left(\alpha-\pi^2\delta_{\ell=2}\right)((Q_1^-)^2 + Q_2^+)\right]}{\prod\limits_{i,j=1}^{J} \left(1-g^2/(x^{+}_i x^{+}_j)\right)\left(1-g^2/(x^{-}_i x^{-}_j)\right)}.
%\eeq

Having expressed the results, a number of remarks are in order. The first is that (\ref{su2structure}) is an exact result capturing finite-volume effects around the seams adjacent to the NBPS operator \cite{bottom}. These effects start at N$^3$LO in the SL(2) sector and therefore were not considered in the previous section, but for SU(2) they are present already at NNLO. Their effect in the SOV representation is encoded in the dressing of the Q-function through the~{\color{blue}{$\delta_{\ell, L-1}$}} factor, see appendix \ref{finitevolume} for a lengthy discussion on these finite volume effects. Here we emphasize that \textit{the structure constant geometry dress the Q-function}: $\mathcal{Q}$ depends both on the state and $\ell$.
\begin{figure}[t]
    \centering
    \includegraphics[width=0.56\textwidth]{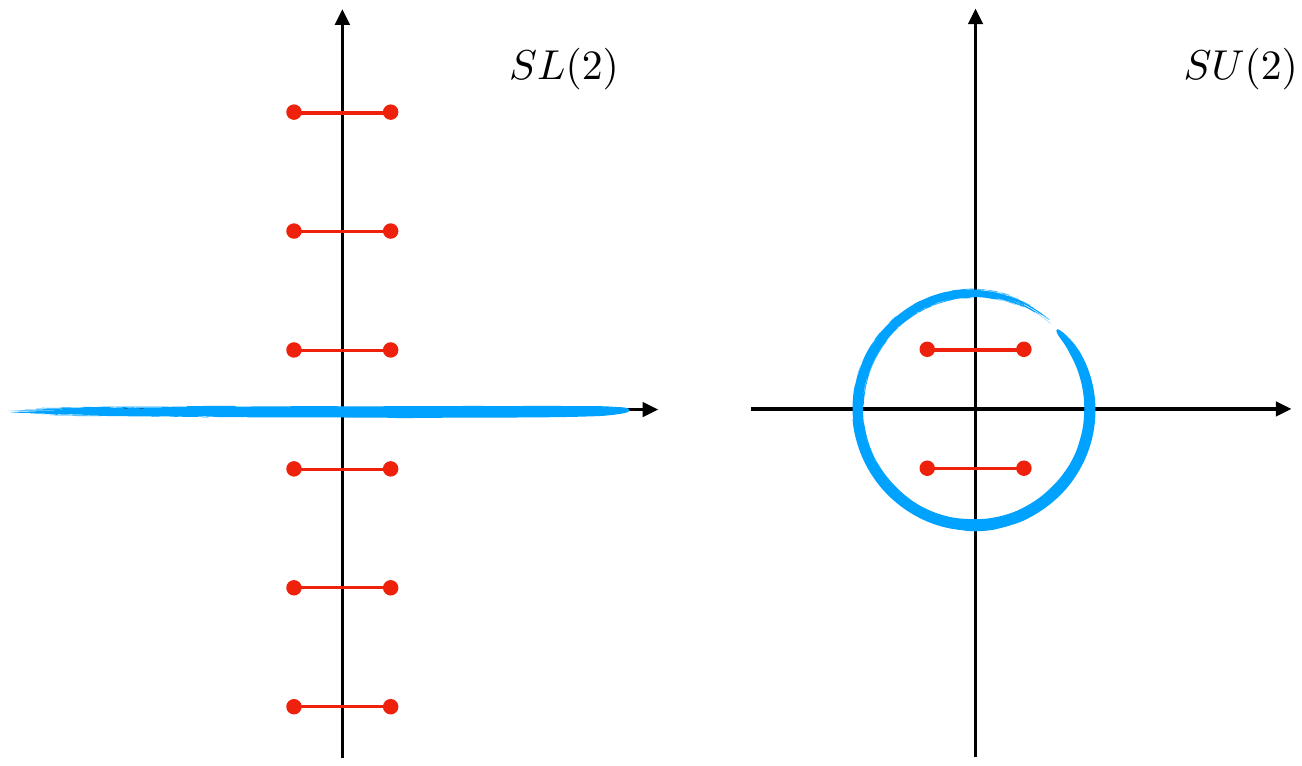}
    \vspace{-2.8cm}
    \caption{For SL(2) we integrate over the real axis; in SU(2) we have a contour integral encircling the Zhukowsky cuts. }
    \label{contours}
\end{figure}

The second remark is the presence of the {\color{red}{$\delta_{\ell=2}$}} ``anomaly" in (\ref{dressing}). Its origin is not clear to us. Should it be put on foot with the $\ell=1$ result (\ref{l1resultsl2}) in the SL(2) sector at NNLO in which a new integral appears? Is it an indication that we are integrating out a simpler higher-dimensional integral for this short bridge overlap? %Next, note that uniform transcendentality is not manifest in (\ref{su2structure}) at NNLO. It would be desirable to identify a change of variables that restore naive transcendentality counting. The correct prescription may also shed light on the $\ell=2$ anomaly. {\color{red}{AH: I'm not happy with the flow in this transcendentality part, please someone refine if they have a suggestion}}

To a more basic point, in contrast to the SL(2) sector result (\ref{AbelianOneSpin}), different measures and Q-dressings enter the numerator and denominator of (\ref{su2structure}) already at NLO. From the $\theta$-morphism point of view, the mismatch is due to the boundary terms in the second line of (\ref{morphismeq}) which cancel when acting on the denominator, a symmetric function of the $\theta_i$, but do not on the numerator. It turns out that the denominator's \textit{Gaudin} measure follows as in the SL(2) case from an orthogonality principle -- see appendix (\ref{su2orthogonality}). The numerator measure is more complicated and we could not identify an orthogonality principle that fixes it. Is there an alternative principle that generalizes to higher loops and allow us to move forward? We hope so. As usual, one can always rely on hexagons to compare any new proposal with data, as we did to confirm the correctness of (\ref{su2structure}).

%{\color{red} P: Maybe this last paragraph should be the first one and be closer to (\ref{dressing})?}

%Finally, note that uniform transcendentality is not manifest in (\ref{su2structure}) at NNLO. 

\section{SL(2) vs SU(2)} \label{comparison}

It is amusing to compare our results so far in the following summary table:

\begin{table}[h!]
\centering
\footnotesize
\begin{tabular}{||c c c||} 
 \hline
& SL(2) & SU(2) \\ [0.5ex] 
 \hline 
 \hline
  Main result & (\ref{AbelianOneSpin}) and (\ref{AbelianThreeSpins})  & (\ref{su2structure}) \\ 
  \hline
 $\begin{array}{c} \text{Factorized} \\ \text{measure} \end{array}$ & yes  & yes \\
 \hline
  $\begin{array}{c} \text{Same $\mu$ and $\mathcal{Q}$} \\ \text{for $\mathcal{A}$ and $\mathcal{B}$} \end{array}$ & yes  & no \\
 \hline
 Contour & Real axis 
 %$\begin{array}{c} \text{Real} \\ \text{axis} \end{array}$
 & $\begin{array}{c} \text{Encircling} \\ x^{\pm}(u) \text{ cuts} \end{array}$ \\
 \hline
 $\begin{array}{c} \text{Manifest} \\ \text{transcendentality} \end{array}$ & yes & no \\
 \hline
 $\begin{array}{c} \text{Wrapping effects} \\ \text{incorporated} \end{array}$ & Bottom: e.g. (\ref{botWrap})  & Adjacent: e.g. (\ref{dressing})  \\
 \hline
$\begin{array}{c} \text{Derivation and} \\ \text{guesswork tools} \end{array}$
  & $\begin{array}{c} \text{Hahn polynomials} \\ \text{Baxter orthogonality} \\ \text{Zhukowskization} \\ \text{Hexagon data} \\ \text{Bottom Wrapping}  \\ \text{Reciprocity}
  \end{array}$ & 
  $\begin{array}{c} \text{$\theta$-morphism} \\ \text{Baxter} \\ \text{Hexagon data} \\ \text{Adj. Wrapping} \\ \text{Selection rules}
  \end{array}$ \\ \hline
 Next step & 
 
 $\begin{array}{c} \text{Finish NNLO} \\ \text{(\& guess all loops)} \end{array}$  &  $\begin{array}{c} \text{Do NNNLO} \\ \text{(\& guess all loops)} \end{array}$  \\ \hline
\end{tabular}
\label{alphaResultTable}
%\caption{Add}
\end{table}

Many things are common to both SU(2) and SL(2) correlators most notably both are given by a SoV like scalar products involving factorized quantum corrected measures and quantum dressed Baxter functions. As far as we checked, both are capable of capturing wrapping corrections. 

There are also differences. Some seem minor: for example, the counting of transcendentality in the SU(2) expressions is a bit weird specially due to the mixed transcendentality factor $\alpha = \tfrac{2}{3} \pi^2 -1$ \cite{transcendentality,transcendentality2}. 

Some differences seem deeper: The contour of the non-compact SL(2) sector is non-compact while the contour for the SU(2) compact sector is a closed contour, see figure \ref{contours}. (This was already observed before in rational SoV explorations, see e.g. \cite{clustering, kolyasu2,kolyasl3}). 

For states with two particles of opposite momenta~$\{v,-v\}$, SL(2) three-point functions of length $L$ and adjacent bride $\ell$ should match SU(2) correlators of length $L+2$ and bridge $\ell+1$. We checked this by explicit evaluation but it is amusing that this type of relation -- which follows from some SU(2) and SL(2) states being in the same supermultiplet -- is manifest in the spectrum problem \cite{BeisertStaudacher} but is totally obscure here. Would be desirable to make it more manifest; this might hint at an even more unified description of both sectors. 

Let us conclude this section highlighting the beautiful appendix G of \cite{clustering} by Jiang, Komatsu, Kostov and Serban. There the starting point are the hexagon sums over partitions for $\mathcal{A}$. These sums are cast as contour integrals and after several cleaver manipulations these contours end up being recast as new SoV like integrals. They do this for both SU(2) and SL(2). (The SL(2) derivation is more involved with some "straightforward (but complicated and tedious)" steps.) In other words, they \textit{derive} SoV from hexagons. The drawback if that this derivation was done at tree level and there are a few steps (see previous quote) that don't seem to be obvious to lift to all loops. 
Revisiting this appendix in light of what we learned seems promising.

\section{Discussion} \label{discussion}
We have initiated here a project for recasting correlators in $\mathcal{N}=4$ SYM in a language closer to that of the spectrum quantum spectral curve. With all recent SoV related advances \cite{giombi,caetano,kolyasu2,kolyasl3,kolyadeterminant,kolyafishnet,Giombi:2018qox,Giombi:2018hsx,Gromov:2015dfa} we believe the time is ripe for seeking out for such a new approach. We are probably scratching the tip of an iceberg as far as the full finite coupling structure goes but some elements are clearly emerging. 

One is the central importance of SoV quantum corrected measures $\mu(u_1,\dots,u_{p})$ which can be found by combining a myriad of different approaches from orthogonality of quantum corrected Baxter equations, matching SoV integrals with hexagon combinatorics, converting hexagons sums into contour integrals which one then massages into SoV like integrals, $\theta-$morphism operations on in-homogeneous SoV integrals, re-summing wrapping corrections, extending orthogonality relations of known classical polynomials, Zhukowsky upgrades of trigonometric functions \textit{etcetera}. (This \textit{etcetera} often includes a good deal of \textit{inspired guesswork}.)

These measures are then used to build scalar product like integrals which couple Baxter functions $\mathcal{Q}(u)$ for the various involved operators. The Baxter functions also are corrected away from their tree level polynomial forms. 

The most obvious question is then how to fix the measures and the Baxter Q-functions. Do these measures obey any sort of all loop bootstrap set of axioms? They must. It is crucial to work it out. Are these Q-functions some of the finite coupling solutions to the Quantum Spectral Curve? We hope so. 

\begin{figure}[t]
\centering
\includegraphics[width=0.5\textwidth]{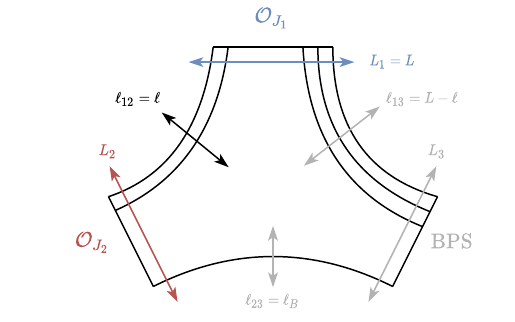}
\caption{The NBPS operators have twists $L_i$ and spins $J_i$. Their polarizations are orthogonal. In the conventions of \cite{spinninghexagons} it is equivalent to $\sum_{\ell}C^{\bullet\bullet\circ}(J_1,J_2,\ell)$.}
\label{Abelian1}
\end{figure}

For the most part we considered a single non-trivial non-BPS operator but we could have considered correlation functions involving \textcolor{black}{more than one} non-BPS operators using the very same universal measures and Baxter functions. Take for the example the case of \textcolor{black}{two} spinning operators as studied in \cite{spinninghexagons} and consider the so-called abelian polarization there, see figure \ref{Abelian1}. We found a compact representation of all such correlators at NLO as 
\beq
\!\! \Big(\sum_{\ell} C^{\bullet\bullet\circ}_{\ell}\Big)^{\!2} \!= \frac{(\textcolor{blue}{\mathbb{J}_1}+\textcolor{red}{\mathbb{J}_2})!^2}{(2\textcolor{blue}{\mathbb{J}_1})! (2\textcolor{red}{\mathbb{J}_2})!} \frac{\langle\textcolor{blue}{\mathcal{Q}_1},\textcolor{red}{\mathcal{Q}_2}\rangle_\ell^2}{\langle \textcolor{blue}{\mathcal{Q}_1},\textcolor{blue}{\mathcal{Q}_1}\rangle_{\textcolor{blue}{L_1}} \langle \textcolor{red}{\mathcal{Q}_2},\textcolor{red}{\mathcal{Q}_2}\rangle_{\textcolor{red}{L_2}}} \,. \label{AbelianThreeSpins}
\eeq

Would be very interesting to find an SoV representation for the other spinning correlator tensor structures studied in \cite{spinninghexagons}; a strategy could be to embed the external operator SL(2)'s into higher rank SL(N) where we could borrow recent SoV technology from \cite{kolyasl3}.

Is the number of SoV integrals going to remain constant or at least grow in a controllable way? And if they grow, can we re-sum the full result into some kind of exotic infinite dimensional sort of scalar products? We have little to say about the last speculative question but as far as the growth of the number of integrals goes, we do seem to find some interesting structure. Take for example the SL(2) correlation function involving a single non-BPS operator with twist $L=2$ and adjacent bridges $\ell=1$. With $\ell=1$ the numerator $\langle \mathcal{Q},\textbf{1}\rangle_1=1$ to LO and NLO since the scalar product in (\ref{scalarProduct}) is an $\ell-1$ dimensional integral. This is of course consistent with the hexagon approach where this numerator is given by the sum over partitions of magnons $\mathcal{A}_1$ weighted by the hexagon form factors -- see (\ref{calAQ}) -- and this sum over partitions normalized as in this paper indeed evaluated to~$1+O(g^4)$. At NNLO $O(g^4)$, however, it is no longer unity. Instead it is given by a cute expression involving Bernouli numbers which we summarize in appendix \ref{bernouliAp}. Also, at this loop order, if the bottom bridge $\ell_{23}=\ell_B=1$ there is a new wrapping correction to the structure constant \cite{hexagons}. Both these new effects can be compactly incorporated by simply modifying the trivial scalar product 
\beqa
&&\langle \mathcal{Q},\textbf{1}\rangle_1 \to 1+ \eta g^4 \int\! du\, \nu(u) \la{botWrap}  \frac{\mathbb{Q}(u)}{\mathbb{Q}(i/2)} \label{l1resultsl2} \\ &&\qquad\qquad\qquad + \texttt{simple contact terms}_\eta + O(g^6) \,, \nn
\eeqa
with the various ingredients spelled out in appendix \ref{bernouliAp} and where $\eta=1$ for $\ell_B>1$ (no wrapping effects) and $\eta=2$ for $\ell_B=1$ (wrapping effects included). It is very encouraging to see wrapping and asymptotic effects cast in such unified way. We see it as evidence that the number of integrals in this SoV approach should increase as we go to higher and higher loops but that increase is not directly related to wrapping effects which can be automatically incorporated (as seen in this bottom wrapping example and also in the SU(2) adjacent wrapping example, see appendix \ref{finitevolume}). This growth of the number of SoV integrals is consistent with the picture that at finite coupling we are dealing with three quantum strings with infinitely many degrees of freedom and in the SoV approach the number of integrals is related to the number of such degrees of freedom \cite{kolyafishnet}.

As we go to higher loops, multiple wrapping effects come in at once and we should make contact with \cite{Bassowrap} where $PSU(2,2|4)$ transfer matrices are shown to appear. Again, our hope is that even in those cases, wrapping and no-wrapping are all cast in the same unified way. 

%Finally we have strong coupling. Obviously 
We should fit strong coupling in this framework. For large operators there could be interesting semi-classical limits \cite{tunneling,shotastrong,Kostov,shotamirror} where one might be able to make progress; for small operators we know very little about what to expect for the Baxter functions and we have nothing intelligent to say. A good starting point there might be \cite{BassoDeLiang} where important sets of wrapping corrections were beautifully resummed for structure constants involving one small operator and two large operators.  

Another limit worth exploring is large spin. This is a particular limit where SoV should be quite powerful since the number of integrals does not grow with spin. Very interesting structures seem to emerge \cite{WIPwithBen} which might also shed light on WL/correlator dualities \cite{classicalPaper1,classicalPaper2} along the lines of the recent works \cite{ourRecentPapers,ourRecentPapers2}. 

It would also be very interesting to think about how all this fits into a higher point function picture \cite{hexagon4point}. What is the SoV description of a four point function of BPS operators? Does it involve integrating over intermediate Baxter functions? Will the measures derived here play a role in this integration? A starting point for these explorations could be recasting the Coronado's all loop octagon correlator \cite{Frank1,Frank2} in this language, perhaps using its determinant representations \cite{DeterminantPapers}. The octagon is a large R-charge correlator though so at the same time we should probably figure out what sort of simplifications take place on the SoV side when we consider large operators. (Another motivation for studying this limit.) 
%Incidentally, working out such simplifications should also be useful to other problems such as making contact with the strong coupling classical limit so there is certainly plenty of motivation for attacking this interesting direction.

We explored the SU(2) and SL(2) sectors of~$\mathcal{N}=4$ SYM. There is life beyond it. We should look for it.

{\begin{center}{\textbf{ACKNOWLEDGMENTS}} \end{center}}
%\section{Acknowledgments}
We thank B.~Basso, J.~Caetano, F.~Coronado, V.~Goncalves, N.~Gromov, V.~Kazakov, S.~Komatsu and E.~Olivucci for numerous enlightening discussions. We are specially grateful to Benjamin Basso for numerous suggestions, and collaboration on several topics discussed here. 
Research at the Perimeter Institute is supported in part by the Government of Canada through NSERC and by the Province of Ontario through MRI. This work was additionally supported by a grant from the Simons Foundation (PV: \#488661) and FAPESP grant 2016/01343-7 and 2017/03303-1. The work of C.B. was supported by the Sao Paulo Research Foundation (FAPESP) under Grant No. 2018/25180-5.

\appendix
\setcounter{secnumdepth}{2} % only chapter and sections will be numbered
\setcounter{tocdepth}{3}  
\tableofcontents

\section{Notation} \label{appendixNot} \label{notation}

\subsection{Roots and Charges} \label{rootscharges}
%The Bethe roots $v_k$ are the zeroes of the Baxter function. 
%At weak coupling 
Up to at least 4 loops, the Bethe roots $v_k$ are the solutions to the asymptotic Bethe equations~\cite{betheequationsref,BeisertStaudacher}
\beq
1=\left(\frac{x_k^{+}}{x_k^-}\right)^L \prod_{j\neq k} \left(\frac{x_j^--x_k^+}{x_j^+-x_k^-}\right)^\eta\frac{1-g^2/(x_j^-x_k^+)}{1-g^2/(x_j^+x_k^-)} \sigma(v_k,v_j)^2 \nn
\eeq
%{\color{red}P: (We can write in Baxter form but is this the best? I had in mind product form but maybe Baxter is better; never thought about it. Will be painful to introduce $\sigma$ in this form I fear)} 
%\beq
%\frac{(x^{+})^L}{B}Q^{++}+\frac{(x^{-})^L}{B}Q^{--} = t Q
%\eeq
for the simplest SU(2) and SL(2) rank one sectors where $\eta=\pm1$ respectively. Here the Zhukowsky variables 
%(which start at tree-level) %{\color{red}{AH: add $x^{[\pm a]}$ definition here}}
\beq
x_k^\pm = x(v_k\pm \tfrac{i}{2}) \,\,\, \text{with}  \,\,\, x(u) =\frac{u\!+\!\sqrt{u^2-4g^2}}{2} \simeq u-\frac{g^2}{u}+\dots \nn
\eeq
where $g^2=\lambda/(4\pi)^2$ is the coupling. (For three point functions $g^{2k}$ effects are $N^kLO$ effects while for the quantum anomalous dimensions they are $N^{k-1}LO$ effects.)
We sometimes also use $x^{[\pm a]} \equiv x(u\pm i a/2)$.
Finally the dressing phase starts only at very high loop order, $\sigma=1+O(g^6)$ and we will not use its explicit expression in this paper. Generating Bethe roots for SL(2) is simple since they are real; we can use an extension of the code in \cite{nordita} for instance; for SU(2) we used \cite{dymasolver} to produce tree level solutions and then we found the loop corrections by linearizing Bethe equations around these seed values.
%\beq
%\sigma=  \text{expansion with charges} \,.
%\eeq

%We use $v_k$ in the text for the Bethe roots and reserve~$u_n$ for separation of variables type integration variables. 
We also define the auxiliary real functions
%\begin{align}
%    q^{+}_k(u)=\frac{i^k}{(x^{+}(u))^k}+(-1)^k\frac{i^k}{(x^{-}(u))^k} \label{qplus}
%    \\
%    q^{-}_k(u)=\frac{i^{k-1}}{(x^{+}(u))^k}-(-1)^k\frac{i^{k-1}}{(x^{-}(u))^k}\label{qminus}
%\end{align}
\begin{align}
&    q^{+}_k(u)=i^{k+0} (x^{+}(u))^{-k}+(-i)^{k+0} (x^{-}(u))^{-k} \label{qplus}
\,,    \\
&    q^{-}_k(u)=i^{k-1} (x^{+}(u))^{-k}+(-i)^{k-1} (x^{-}(u))^{-k}\label{qminus} \,,
\end{align}
which can be used to construct the conserved charges
\begin{equation}
    Q^{\pm}_k =\sum_{i=1}^J q^{\pm}_k(v_j) \,.
\end{equation}
The $Q^{+}_k$ are even and $Q^{-}_k$ are odd functions of the Bethe roots. The anomalous dimension of an operator is nothing by $\gamma=2g^2 Q_1^+$ for instance. 

We can also package the Bethe roots into Baxter polynomials $\mathbb{Q}(u)\equiv \prod_{k=1}^J (u-v_k)$ or dressed Baxter functions as in (\ref{baxterDef}) or (\ref{dressing}). Then these charges can be extracted by simple contour integrals as well. For instance:
\beq
Q_1^+= \oint \frac{ u\, du}{4\pi g^2 \sqrt{u^2-4g^2}} \log \frac{\mathbb{Q}(u+i/2)}{\mathbb{Q}(u-i/2)}
\eeq
where the contour encircles the Zhukowsky cut~$u \in [-2g,2g]$. This is an interesting definition since it now applies to any function $\mathbb{Q}$, be it a polynomial or not. We can use such definitions to 
%make sense of the scalar products such as (\ref{scalarProduct}) in an off-shell sense to be able to 
pair up arbitrary functions in scalar products such as (\ref{scalarProduct}). 

%In $\mathcal{N}=4$ SYM 
The concept of transcendentality is often  useful. At loop order $n$ we expect functions of uniform transcendentality $n$ to show up once properly counted. It is tempting to assign transcendentality $k$ to the charges $q_k^\pm$. Some identities like~$(q_1^+)^2-(q_1^-)^2 = 2q_2^+$ nicely preserve this counting but others such as $(q_1^+)^2 + (q_1^-)^2 = 4 q_1^+ +O(g^2)$  can lead to ambiguities in perturbation theory. In the $SL(2)$ sector Harmonic numbers $H_k$ of transcendentality $k$ often show up. We use the \texttt{mathematica} notation $H_k(x)=\texttt{HarmonicNumber}(x,k)$ and
%. The following combinations are also quite useful:
\beqa
H^{+}_n(u) &\equiv& H_n(-1/2+i u)+H_n(-1/2-i u)  \,,\\ i H^{-}_n(u)&\equiv& H_n(-1/2+i u)-H_n(-1/2-i u) \,.
\eeqa

\subsection{Structure Constants and Hexagons}

\begin{figure}
\centering
\includegraphics[width=0.45\textwidth]{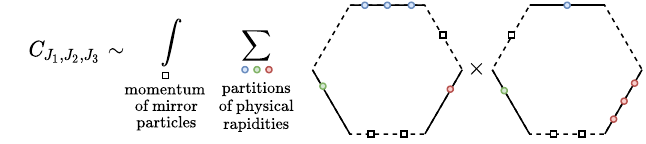}
\caption{Each closed spin chain operator is split into two open chain operators. We sum over all the ways its excitations can end up in either one of the chains. Gluing the hexagons together amounts to integrating over all possible mirror states.}
\label{addHexagons}
\end{figure}

The hexagon formalism \cite{hexagons} is a non-perturbative integrability framework developed to evaluate structure constants in planar $\mathcal{N}=4$ SYM. It was used throughout this letter to generate a myriad of perturbative data. This formalism entails two main components: asymptotic sums and mirror corrections, as depicted in figure \ref{addHexagons}.

When we cut the three point function pair of pants into two hexagons the excitations on each operator can end up on either hexagon so we must sum over the ways of partitioning such excitations. In the end when we glue back the hexagons into a pair of pants we must sum over all possible quantum states along each edge where we glue. In this appendix we ignore these second effect, focusing only in the asymptotic contributions.

The three point function depicted in figure \ref{lBridge} in the hexagon formalism is given by the ratio of two quantities
\begin{equation}
    \left(C^2_{\bullet\circ\circ}\right)_\ell = \frac{\mathcal{A}^2_\ell}{\mathcal{B}}\,.
    \label{CfromHex}
\end{equation}
The numerator entering the ratio above is the central object of the hexagons approach. In the asymptotic regime
\begin{equation}
    \mathcal{A}_\ell^\texttt{asymptotic} = \mathbb{N}_{J,\ell} \sum_{\alpha \cup \bar{\alpha} = \mathbf{v}}(-1)^{|\bar{\alpha}|}\frac{e^{-i p_{\bar{\alpha}} \ell}}{h_{\bar{\alpha}\alpha }}
    \label{mathcalA}
\end{equation}
where $p$ is the momentum of the excitation
\begin{equation}
    p_{\bar{\alpha}} = \sum_{v_i \in \bar{\alpha}} p(v_i) \quad \text{where} \quad e^{i p(u)} = {x^{+}(u)}/{x^{-}(u)}
\end{equation}
and the so-called dynamical factor $ h_{\bar{\alpha}\alpha } = \prod\limits_{\substack{v_i \in \bar{\alpha}\\
  v_j \in \alpha}} h(v_i,v_j)\,,$
%\begin{equation}
% \end{equation}
%where
%which is distinct for each of the rank one sectors
\begin{equation}
    h(u,v)= \left(\tfrac{x^{-}(u)-x^{+}(v)}{x^{+}(u)-x^{-}(v)}\right)^{\eta}\tfrac{x^{-}(u)-x^{-}(v)}{x^{-}(u)-x^{+}(v)}\tfrac{1-\tfrac{1}{x^{-}(u)x^{+}(v)}}{1-\tfrac{1}{x^{+}(u)x^{+}(v)}}\tfrac{1}{\sigma(u,v)} 
    \label{dynamicalfactor}
\end{equation}
with $\eta=0$ for SL(2) and $\eta=1$ for SU(2).

The denominator of (\ref{CfromHex}) is the normalization of the three point function by its two point function constituents
\begin{equation}
    \mathcal{B} = \frac{|\mathbb{N}_{J}|^2 \text{det}(\partial_{v_i}\phi_j)}{\prod_{i=1}^{J}\mu(v_i)\prod_{i\neq j}^Jh(v_i,v_j)}
\end{equation}
where $e^{i\phi_j} \equiv e^{ip(v_j)L}\prod_{k\neq j}S(v_j,v_k) $ and  
%\begin{align}
%&    e^{i\phi_j} \equiv e^{ip(v_j)L}\prod_{k\neq j}S(v_j,v_k) \,,\\
    %\nu(u) &= \frac{\left(1-\frac{g^2}{x^{+}(u)x^{-}(u)}\right)^2}{\left(1-\frac{g^2}{(x^{+}(u))^2}\right)\left(1-\frac{g^2}{(x^{-}(u))^2}\right)}\,.\\
    %\mu(u) &= \left(1-\frac{g^2}{x^{+}x^{-}}\right)^{\!2}\!\left(1-\frac{g^2}{(x^{+})^2}\right)^{\!-1}\!\left(1-\frac{g^2}{(x^{-})^2}\right)^{\!-1}\, \nn
%\end{align}
\begin{align}
%&    e^{i\phi_j} \equiv e^{ip(v_j)L}\prod_{k\neq j}S(v_j,v_k) \,,\\
    %\nu(u) &= \frac{\left(1-\frac{g^2}{x^{+}(u)x^{-}(u)}\right)^2}{\left(1-\frac{g^2}{(x^{+}(u))^2}\right)\left(1-\frac{g^2}{(x^{-}(u))^2}\right)}\,.\\
    \mu(u) \!=\! \left(1\!-\!g^2/(x^{+}x^{-})\right)^{\!2}\!\left(1\!-\!g^2/(x^{+})^2\right)^{\!-1}\!\left(1\!-\!g^2/(x^{-})^2\right)^{\!-1}\, \nn
\end{align}
is the hexagon measure, not to be confused with the various SoV measures. The factor $|\mathbb{N}_J^2|$ is the ($\ell$-independent) absolute value of $\mathbb{N}_{J,\ell}$.
The normalization cancels when evaluating the physical three point function but matters when comparing with the SoV formalism. It is given by
\begin{align}
    \mathbb{N}_{J,\ell} &= \Big(\prod\limits_{i,j}\big( 1-\frac{g^2}{x_i^{+}x_j^{-}}\big)\Big)^{-\eta}\binom{\mathbb{J}}{\mathbb{J}-J}^{\eta-1}\times\frac{i^{2\ell+J+1}}{J!}\times\nonumber\\
    &\times\prod_{i=1}^{J} \sqrt{x^{+}_ix^{-}_i}\times e^{-\frac{i}{2}p(v_i)\ell}
\end{align}
where  $\eta=0$ for SL(2) and $\eta=1$ is for SU(2).

Similar to the hexagon formalism, our SoV expressions for the three point functions are also given by the ratio of two quantities. We were able to directly match each one of the inner products entering the SoV with its hexagon formalism counter parts. For SL(2) it is simply
\begin{equation}
\mathcal{A}_\ell =\langle \mathcal{Q},\mathbf{1} \rangle_\ell \la{calAQ} \,, \qquad \mathcal{B} = \frac{(2\mathbb{J})!}{(\mathbb{J}!)^2}\langle \mathcal{Q},\mathcal{Q} \rangle_L \,,\,\,\,\,\, 
\end{equation}
%\begin{equation}
%\mathcal{B} = \frac{(2\mathbb{J})!}{(\mathbb{J}!)^2}\langle \mathcal{Q},\mathcal{Q} \rangle_L
%\end{equation}
and for SU(2) it reads
\begin{equation}
    \mathcal{A}_\ell = \Lambda_\mathcal{A} \llangle \mathcal{Q},\mathbf{1}\rrangle_{\ell,L}
%\end{equation}
%\begin{equation}
\,,\,\,\,\,\,    \mathcal{B} = \Lambda_\mathcal{B} \frac{(2J)!}{(J!)^2} \llangle \mathcal{Q},\mathcal{Q}\rrangle_{L,L}\, \label{finalgaudin}.
\end{equation}
where
\begin{align}
\Lambda_\mathcal{A} &=e^{g^4 \left(\alpha-\textcolor{red}{\delta_{\ell=2}}\pi^2\right)\left((Q^{-}_1)^2+Q^{+}_2 \right)} \label{LambdaB}\\
\Lambda_\mathcal{B} &=\prod\limits_{i,j}^{J}\left( 1-\tfrac{g^2}{x^{+}(v_i)x^{+}(v_j)}\right)\prod\limits_{i,j}^{J}\left( 1-\tfrac{g^2}{x^{-}(v_i)x^{-}(v_j)}\right) \nn
\end{align}
The normalization factor in (\ref{su2structure}) is then $\Lambda_\ell(Q)= \Lambda_\mathcal{A}^2/\Lambda_\mathcal{B}$.

\section{SL(2) material}
\subsection{Hahn Polynomials and Measures}\la{HahnDetails}
A Hahn polynomial $p_J(x|a,b,c,d)$ is given by 
%(up to an irrelevant overall constant)
\beq
\, _3F_2\left(\!\left.\begin{array}{c} a+i x\,,\, a+b+c+d-1+J \,,\, -J\\ 
a+c\, , a+d\end{array} \right|1\right) \la{HahnDef}
\eeq
and admits a simple orthogonality relation
\beq
\int dx \,\mu_\text{Hahn}(x) \,p_J(x|a,b,c,d) p_J'(x|a,b,c,d) \propto \delta_{JJ'} \label{orthoHahn}
\eeq
with 
\beq
\mu_\text{Hahn}(x)= \Gamma(a+i x)\Gamma(b+i x)\Gamma(c-i x)\Gamma(d-i x) \,. \label{Hahnmu}
\eeq
At LO the twist-2 Baxter functions are given by these polynomials with $a,b,c,d=\tfrac{1}{2}$ so that the measure becomes $\Gamma(\tfrac{1}{2}+ix)^2 \Gamma(\tfrac{1}{2}-i x)^2 \propto \text{sech}^2(\pi x)$ as quoted in~(\ref{onePMeasure}). 

At loop level the corrections to $1/2$ for the coefficients $a,b,c,d$ depend on $J$ albeit mildly, through the Harmonic number $H_1(J)$, so that (\ref{HahnDef}) reads 
\beq
\, _3F_2\left(\!\left.\begin{array}{c} \tfrac{1}{2}+ {\color{magenta} i \sqrt{2} g}+{\color{magenta} 2 g^2 H_1}+i {\color{blue} u}\,,\, J+1+{\color{magenta} 8 g^2 H_1}\, , \, -J\\ 
1+{\color{magenta} 4 g^2 H_1}\,,\,1+{\color{magenta} 2 i \sqrt{2} g}+ {\color{magenta}4 g^2 H_1}\end{array} \right|1\right)\nn
%\la{3F2}
\eeq
where we highlighted the NLO corrections in magenta. 
%and the argument of $H_1$ is $J$. 
%That requires us to be careful when using (\ref{orthoHahn}). 
If we plug the $J$ dependent one loop values for $a,b,c,d$ in (\ref{Hahnmu}) we would obtain
\beqa
\mu_\text{Hahn}(x) \propto\frac{1}{\cosh^2(\pi x)} \Big(1 +2\pi^2 g^2 \alpha  \tanh^2(\pi x) \nn \\  + 4 g^2 H_1(J)H_1^+(x) + O(g^4) \Big) \,. \la{muExpanded}
\eeqa
where $\alpha=1$. 
This is unsatisfactory as it is $J$ dependent. To fix this we tried to absorb the second line into the definition of $\mathcal{Q}_J$; recalling that these Harmonic numbers are nothing but the energy $Q_1^+$ of these twist-two operators we end up with the correction in the first line in~(\ref{baxterDef}). (We can not derive the second line with this simple derivation since the odd charge $Q_1^-$ vanishes for the operators in this leading Regge trajectory.) What we did is then absorb this second line of (\ref{muExpanded}) into the Baxter functions and look for an orthogonal measure for these \textit{no longer polynomial} Baxter functions of the form of the first line of (\ref{muExpanded}). It exists with $\alpha=3/2$. In (\ref{onePMeasure}) we fixed the overall normalization of the measure so that the vacuum ($Q_0=1$) is unit normalized. 

%to obtain the expansion in (\ref{onePMeasure}) up to an overall constant. Also, the Harmonic numbers in (\ref{3F2}) are nothing but the energy $Q_1^+$ of these operators and lead to the correction in the first line in (\ref{baxterDef}) for these twist two operators, see appendix \ref{HahnDetails}. We can not derive the second line with this simple derivation since the odd charge $Q_1^-$ vanishes for the operators in this leading Regge trajectory.   

There are still interesting open problems to pursue along these lines. NNLO corrections to twist-two Baxter polynomials are also known \cite{NLOHahn}. Can we use them to infer the next correction to the measure? Some twist three families are also known \cite{twist3,NLOHahn}. Can we use them to shed light on the multi-particle measure $\mu_2$? 

\subsection{SL(2) Baxter and Measure} \label{BaxterSL2Ap}

It is easy \cite{straight} to check that for on-shell solutions of the one-loop twist 2 Baxter equation (\ref{baxterT}) the dressed polynomials (\ref{baxterDef}) solve the simplified difference equation
\beq
\label{simplifiedBaxter} \tilde{B} \circ \mathcal{Q}(u) = T(u) \mathcal{Q}(u) 
\eeq
with $\tilde{B} \equiv (x^+)^2 e^{i\partial_u} + c.c.$ without any charge dependence in contradistinction with the original $\mathcal{B}$. 

From this is easy to build a one loop orthogonal inner product for the twist 2 solutions. We look for a measure so that $\tilde{B}$ is self adjoint in the inner product
\beq
\langle \mathcal{Q}_i, \mathcal{Q}_j \rangle \equiv \int dx\, \mu_1(x) \mathcal{Q}_i\mathcal{Q}_j \label{innersimple}
\eeq
on the space of dressed polynomials (\ref{baxterDef}). Deforming the contours in $\langle \tilde{B} \circ \mathcal{Q}_1, \mathcal{Q}_2 \rangle $ as done in section \ref{SL2comments} shows that it is enough to consider a periodic measure with fast enough decay at infinity provided all poles of the integrand -- which now receive contributions from $\mu_1$, $\mathcal{Q}$ and $\tilde{B}$ -- have vanishing combined residue. See \cite{kolyasl3} for similar ideas applied to SL(N) spin chains. Writing an ansatz
\beq
\mu_1(x) = \frac{\pi/2}{\cosh^2(\pi x)}\left(a_1 + g^2 \[a_2 \tanh^2(\pi u) + a_3\]\right)
\eeq
and requiring the cancelation of poles combined with the requirement that the vacuum ($\mathcal{Q}_0 = 1$) is  normalized to one in (\ref{innersimple}) fix $a_1 = 1$, $a_2 = 3\pi^2$, $a_3 = -1$. This is (\ref{onePMeasure}).

\subsection{Hexagons $\mathcal{A}$ at $\ell=1$}  \la{bernouliAp}
For minimal left bridge length $\ell=1$ the hexagon sum over partitions simplifies dramatically. We thank Frank Coronado for highlighting this and for help in establishing the asymptotic formulae~(\ref{bernouli}) a few years ago. 

The asymptotic result (\ref{mathcalA}) for $\ell=1$ simplifies to
\beq
(\mathcal{A}^\texttt{asymptotic}_{\ell=1})^2 = 1+g^4 \sum_{k=3}^S\label{bernouli}\sum_{i_1<\dots<i_k} \! \frac{i^k (4-k)B_{3-k}}{\prod\limits_{l=1}^k \left(v_{i_l}+\frac{i}{2}\right)}+\,{c.c.} \nn
\eeq
plus $O(g^6)$. Here
%\begin{align}
%(\mathcal{A}^\texttt{asymptotic}_{\ell=1})^2 &= 1+g^4\Bigg( \sum_{k=3}^S(4-k)B_{3-k}\times\label{bernouli}\\
%&\times\sum_{i_1<\dots<i_k}  \frac{i^k}{\prod\limits_{l=1}^k \left(v_{i_l}+\frac{i}{2}\right)}+c.c.\Bigg)+O(g^6) \nn
%\end{align}
%\beqa
%&&(\mathcal{A}^\texttt{asymptotic}_{l=1})^2 =  \frac{e^{2g^2 H_1(S) q_1^+ - g^4 H_{2}(S) (q_1^+)^2}}{\prod\limits_{j=1}^S x_j^+x_j^-}  \times \la{bernouli} \\
%&&\times\Big(1+g^4 \sum_{k=3}^S\frac{4-k}{2}B_{3-k}\!\!\! \sum_{i_1<\dots<i_k}  \frac{i^k}{\prod\limits_{l=1}^k (u_{i_l}+i/2)}+c.c.\Big) \nn
%\eeqa
 $B_{n}$ are the Bernouli numbers.

Since the roots $v_i$ that appear in the denominator product never appear repeated we can cast the sum as
\beq
1+g^4 \frac{P(v_j)}{Q(-i/2)}+ c.c. \la{cc}
\eeq
where $P(v_k)$ is a polynomial \textit{linear} in each of the Bethe roots~$v_j$, just like the Baxter polynomial~$Q(u)\equiv \prod (u-v_k)=\sum_n c_n u^n$ is. We can thus look for a linear integral operator acting on $Q$ and producing such $P(v_j)$,
\beq
\< Q(u) \> = \sum_{n=0}^S c_n(v_j) \<u^n\>  = P(v_j) \,. \label{need}
\eeq
(It is not granted that such operation exists for all $S$.)
%but that turns out to be the case. 
We imposed it spin by spin and observed that we can indeed satisfy (\ref{need}) as long as we fix the linear map moments as
\beq
\<1\> = \< u\>=\<u^2\>=0 \,, \<u^3\>=\tfrac{i}{2} \,, \<u^4\>=1\,, \<v^5\>=-\tfrac{7i}{6} \,, \dots \nn
\eeq
We computed the first few dozens such moments from which we guessed that they can be generated from
\beqa
\<F \> \equiv \tfrac{\pi^2}{15} F(\tfrac{-i}{2})+ 3i \zeta_3 F'(\tfrac{-i}{2})-\tfrac{\pi^4}{12} F''(\tfrac{-i}{2})+\!\!\int\!\! du\,  \nu(u) F(u)\nn
%\la{nu} \,,
\eeqa
where the measure 
\beq
\nu(u)=\frac{i \pi -i(u+\frac{i}{2})\pi^2  \tanh (\pi  u) }{2 \left(u+\frac{i}{2}\right)^3 \cosh^2(\pi u)}\,.
\eeq
This came as a surprise as this very same measure arose before in a very different context. Once we use that $Q(\tfrac{i}{2})=Q(-\tfrac{i}{2})$ for physical states to combine both terms in (\ref{cc}) we see that (the real part of) this measure is nothing but the measure which arose when computing the first wrapping correction when the bottom bridge $\ell_B=1$, see equations (44),(55) in \cite{hexagons}! We  conclude that we can not only cast the two loop asymptotic contribution (\ref{bernouli}) as a simple SoV looking integral but we can also trivially incorporate the bottom wrapping effects:
%, with a few simple factors of $2$:
%. In sum, we find 
%But this is nothing but the wrapping contribution from the bottom edge worked out in (\ref{hexagons})! We thus conclude that the full result for the structre constant -- including the leading bottom wrapping effect -- simply reads
\begin{align}
     &\left(\mathcal{A}^\texttt{asymptotic + bottom wrapping}_{\ell=1}\right)^2 = 1+ \label{bernouli2} \\
     &+g^4\Big( \eta\tfrac{\pi^2}{12}Q_1^{+} - \eta\tfrac{\pi^2}{6}Q_2^{+}-6(2-\eta)\zeta(3)Q_1^{+}+\eta \tfrac{4\pi^4}{15}+\nonumber \\
     & +16\pi\eta \int\tfrac{1-12u^2+2\pi u(1+4u^2)\tanh(\pi u)}{(1+4u^2)^3\cosh^2(\pi u)}\tfrac{Q(u)}{Q(i/2)}\,du\Big)+\dots\nonumber
\end{align}
%\beqa
%&&\!\!\!\!\!\!\! (\mathcal{A}^\texttt{asymptotic + bottom wrapping}_{l=1})^2 = \text{Add binomials here to simplify next line}
%\label{bernouli2} \\
%&&=\frac{e^{2g^2 (H_1(S) -3 \eta \zeta_3 g^2) q_1^+ - \frac{g^4}{2}(H_{2}(S)-\eta \frac{\pi^2}{12}) (q_1^+)^2-\eta \frac{\pi^2}{24} g^4 (q_1^-)^2 + \eta \frac{2\pi^4}{15}g^4 \dots }}{\prod\limits_{j=1}^S x_j^+x_j^-}  \times \nn  \\
%&&\times\Big(1- g^4 \eta \int \frac{1-12 u^2+ 2\pi (1+4u^2)  \tanh (\pi  u) }{\tfrac{1}{4\pi^2} \left(4u^2+1\right)^3 \cosh^2(\pi u)} \frac{Q(u)}{Q(i/2)} \Big) \,, \nonumber
%\eeqa
where $\eta=1$ for $\ell_B>1$ and $\eta=2$ for $\ell_B=1$ \cite{techremark}.

Do we continue to have identical formulas with and without wrapping at higher loops? The NNNLO effective wrapping measure correction was nicely worked out in appendix B of \cite{wrappingthreeES} (in appendix E of \cite{wrappingthreeBGKV} an equivalent representation -- a sort of Fourier transform -- was derived). Can the next loop order asymptotic result still be neatly combined with the wrapping correction? 

At higher loops we will get more mirror corrections which we might be able to cast as SoV like integrals. We should probably expect a similar growth in the number of integrals for the asymptotic part of the result as well. Ultimately, at finite coupling, the distinction between the two should fade away as in the spectrum problem. 

\section{SU(2) material}
\subsection{$\theta$-morphism at NNLO}

\label{morphismappendix}
The morphism operator $\mathcal{M}$ performs a ``Zhukowskization" of the rational propagation of magnons in the XXX$_{1/2}$ through the action on background inhomogeneities.
From our perspective $\mathcal{M}$ is defined by equation (\ref{morphismaction}). We write an ansatz and fix it by requiring the match of the RHS of (\ref{morphismaction}) with $C^2_{\bullet \circ \circ}$ which can be computed from hexagons. It can be divided into a \textit{closed} part and a \textit{boundary} part,
\beq
\mathcal{M} = \mathcal{M}_\text{c} \cdot \mathcal{M}_\text{b} + O(g^6). \label{morphisminappendix}
\eeq
At NLO we recover the result in \cite{morphism}; at NNLO we obtain
\beq
\mathcal{M}_\text{c} = \exp\Big(   \sum_{i=1}^L  \Big(g^2  (\partial_{i, i+1})^2  - \frac{1}{4} g^4  (\partial_{i,i+1})^2(\partial_{i+1,i+2})^2\Big) \Big)\,, \nonumber
\eeq
and
\beq
\mathcal{M}_b = \exp  \big( -i g^2  Q^+_1 (\partial_{1}  - \partial_{L}) + g^4 \delta \mathcal{M}_\text{NNLO-b}\big), \label{boundary}
\eeq
with
\begin{align}
&\mathcal{M}_\text{NNLO-b} = \tfrac{1}{2} \left( 2 (Q^+_1)^2 -i Q^-_2 - Q^-_1 Q^+_1\right) \nn \partial_1^2 + i Q^+_1\partial_1^3 + \\& \nn \tfrac{1}{2}  \left(Q^-_1 Q^+_1 - (Q^+_1)^2 \right) \partial_2^2 - \tfrac{1}{2} i Q^+_1 \partial_2^3 - 
    \tfrac{i}{2}   Q^+_1 \partial_1 \partial_2^2 +\\& \tfrac{1}{2} (i Q^-_2 -(Q^+_1)^2) \partial_1 \partial_2 - \left(\partial_1 \leftrightarrow \partial_L, \partial_2\leftrightarrow \partial_{L-1} \right).
\end{align}
It would be interesting to re-derive $\mathcal{M}_c$ and $\mathcal{M}_b$ along the lines of \cite{morphism}.

The \textit{closed} action satisfies the \textit{morphism} property when acting on symmetric functions $f_\text{sym}(\theta)$ of the inhomogeneities $\theta_j$:
\beq
\mathcal{M}_\text{c}  \circ \left(f_\text{sym} \, g \right) = \left(\mathcal{M}_\text{c}  \circ f_\text{sym}\right) \left(\mathcal{M}_\text{c} \circ g \right)  \label{morphismproperty}
\eeq
for a generic function $g$. It also satisfies the ``Zhukowskization" property
\beq
\mathcal{M}_\text{c} \circ \prod_{i=1}^L \left(u - \theta_i \pm i/2\right)^k = \left(x^\pm\right)^{k L} \la{xproperty}.
\eeq
The inhomogeneous Gaudin norm (\ref{gaudinwiththeta}) is given in the SOV representation, see \cite{shotaXXX} for details, by 
\beq
\mathcal{B}_\theta = f_\text{sym}(\theta) \times \oint d\tilde{\mu}  \mu_\theta \prod_{j=1}^{L-1} \mathcal{Q}(u_j)^2
\eeq
with the $\theta$ dependence entering only through the symmetric normalization function $f_\text{sym}(\theta)$ and the measure 
\beq
 \mu_\theta = \prod_{i=1}^{L-1}\prod_{j=1}^L ((u_i - \theta_j)^2 + 1/4)^{-1}.
\eeq
Combining (\ref{morphismproperty}), (\ref{xproperty}) we get the NNLO Gaudin measure
\beq
d\mu_{L,L} = d\tilde{\mu} \times \left(\mathcal{M}_c \circ \mu_\theta = (x^+_j x^-_j)^{-L}\right)
\eeq
as in (\ref{periodicmeasuresu2},\ref{measure1su2},\ref{measure2su2}). The same measure is derived from a orthogonality principle in section (\ref{su2orthogonality}). Comparison with the NNLO gaudin determinant fixes the conversion  factor $\Lambda_\mathcal{B}$ to (\ref{LambdaB}), so that in the end we are left with 
\beq
\nn |\Lambda_\mathcal{B}  \mathcal{M}\circ \mathcal{B}_{\theta}| = \mathcal{B},
\eeq with $\mathcal{B}$ given in (\ref{finalgaudin}) \cite{massage}. Note that $\mathcal{M}_b$, (\ref{boundary}), acts trivialy in $\mathcal{B}_\theta$ since it is a symmetric function of $\theta_i$.

Having fixed the morphism operator through hexagons, the SOV result (\ref{su2structure}) then follows \cite{massage}  from the action of the morphism operator (\ref{morphisminappendix}) on the inhomogeneous XXX$_{1/2}$ 
spin chain SOV overlaps~\cite{shotaXXX,clustering,kolyasu2}. 

\subsection{SU(2) orthogonality}
\label{su2orthogonality}

The \textit{Gaudin} measure defining $\langle \mathcal{Q}, \mathcal{Q} \rangle_{L,L}$ was derived in sections \ref{SU2comments}, \ref{morphismappendix} from the $\theta$-morphism action. Here we show it also defines an orthogonal scalar product for NNLO Baxter polynomials $\mathbb{Q}$, meaning
\beq
\langle \mathbb{Q}_1, \mathbb{Q}_2 \rangle_{L,L} = \mathcal{N}_1 \delta_{12} 
\eeq
if $\mathbb{Q}_1, \mathbb{Q}_2$ are solutions to the Baxter equation $\mathcal{B} \circ \mathbb{Q} = T(u) \mathbb{Q}(u)$ with
\beq
\mathcal{B} = (x^+)^L e^{- i \partial_u} + (x^-)^L e^{ i \partial_u} \,.
\eeq
What follows is a simple loop generalization of the XXX case \cite{kolyasu2}. Consider the pairing 
\beq
\langle f,g\rangle_\mu \equiv\oint_\gamma du \,\mu(u) f(u) g(u) \label{su2orthopair}
\eeq
with the $\gamma$ countour being the boundary of the $[-3g, 3g] \times [-2i,2i]$ square. 
Inserting $\mathcal{B}$,
\beq
\langle \left((x^+)^L e^{- i \partial_u} + (x^-)^L e^{ i \partial_u}\right)\circ f,g\rangle_\mu ,
\eeq
and shifting contours down by $i$ in the first term and up by $i$  in the second term, as done below equation (\ref{baxterDerivation}), we find that $\mathcal{B}$ is self-adjoint with respect to (\ref{su2orthopair}) provided 
%\beq
%\frac{\mu^{[2]}}{\mu} = \left(\frac{x^{[-1]}}{x^{[3]}}\right)^L,
%\eeq
\beq
{\mu^{[2]}}/{\mu} = ({x^{[-1]}}/{x^{[3]}})^L,
\eeq
so that $\mu ={(x^+ x^-)^{-L}} \mu_\text{p}  $ with $\mu_\text{p}$ an~$i-$periodic factor.

Note that $T(u)$ is polynomial for physical zero-momentum states at NNLO  
\beq
T(u) = 2 u^L + \sum_{i=0}^{L-2} c_i(\mathbb{Q}) u^i
\eeq
with $c_i(\mathbb{Q})$ being the state-dependent integrals of motion. Consider the family of measures
\beq
\mu_j = \frac{\sinh(2\pi u)}{(x^+ x^-)^L}\times \exp[2 \pi u (2 j-L)]
\eeq
with $j = 1,\dots,L-1$. Let $\mathbb{Q}_{1},\mathbb{Q}_{2}$ be solutions of the Baxter equation. We then have, from self-adjointness,
\beq
\sum_{i=0}^{L-2} (c_i(\mathbb{Q}_1)- c_i(\mathbb{Q}_2)) \langle \mathbb{Q}_1, u^i \mathbb{Q}_2\rangle_{\mu_j} =0 \label{linearsystem}
\eeq
since the LHS is simply $ \langle \mathcal{B}\circ\mathbb{Q}_1, \mathbb{Q}_2\rangle_{\mu_j}-\langle \mathbb{Q}_1, \mathcal{B} \circ \mathbb{Q}_2\rangle_{\mu_j}=0$. The integrals of motion $c_i$ are generic and therefore the linear system (\ref{linearsystem}) should be non degenerate, implying
\beq
\det\left[ \langle \mathbb{Q}_1, u^{i-1} \mathbb{Q}_2\rangle_{\mu_j}\right] =  \mathcal{N}_1 \delta_{12} . \label{detsu2gaudin}
\eeq
with $i,j = 1,\dots,L-1$. Our claim is that expanding the Vandermonde-like determinant (\ref{detsu2gaudin})into a $L-1$ dimensional integral reproduces the main text result (\ref{periodicmeasuresu2},\ref{measure1su2},\ref{measure2su2}) for $\ell = L$ up to a combinatorial normalization factor,
\beq
\det\left[ \langle \mathbb{Q}_1, u^{i-1} \mathbb{Q}_2\rangle_{\mu_j}\right] \propto \oint\limits_\gamma d\mu_{L, L}  \prod_{i=1}^{L-1} \mathcal{Q}_1(u_i) \mathcal{Q}_2(u_i). \nonumber
\eeq
We conclude that the \textit{Gaudin} measure defines and orthogonal scalar product and, more over, the \textit{Gaudin} norm takes determinant form in the SOV representation.

For $\mathcal{A}$ things are more involved as explained in the main text with the additional exponential dressings in~(\ref{measure1su2},\ref{measure2su2}) kicking in. 

\subsection{SU(2) structure constants at finite volume}
\label{finitevolume}
%In this letter we consider the decay of an excited string into two protected states. In the finite geometry of the pair of pants describing such process, virtual mirror particles might nucleate and interact with the physical excitations describing the non-protected operator before being reabsorbed.

At leading order mirror particles have infinite energy and vanishing phase space, and therefore can be ignored.  Their contribution, which starts at NNLO, is however crucial. It is only when they are taken into account that selection rules are realized for instance. 
In this appendix, we review how the first mirror contributions are computed in the hexagon formalism, adapting the SL(2) computations of \cite{wrappingthreeBGKV,wrappingthreeES,wrappingshota} to the SU(2) case. 
%We refer to these references for details on finite volume corrections in the hexagon formalism. 
We then discuss how these virtual effects are taken into account in the SOV approach by appropriately dressing the Q-functions.
%- and how this dressing was determined.

{\begin{center}{\textit{Part I: mirror particles on the hexagon}} \end{center}}
%\section{Acknowledgments}
\label{Part1}
When gluing hexagons together to reconstruct the closed string geometry one must insert a complete basis of states along the seams $ (\ell_{12},\ell_{31},\ell_{23})$. This complete set of states is given by the Fock space of mirror particles labeled by their rapidities $v$ and bound state index $a$.
The contribution from terms with $n_{ij}$ particles on edge $\ell_{ij}$ is 
%given by
\beq
\mathcal{A}(n_{ij}) = \sum_{\{\textbf{a}\}}  \int d\mu(\textbf{v})e^{-\tilde{E}(\textbf{v}_{ij})\ell_{ij}} \mathcal{H}^2(\textbf{v},\textbf{a}) \nonumber
\eeq
where $\{\textbf{a}\} = \{a_k^{ij}\}$ is a collective index for the bound state label $a_k^{ij}$ of particle $k$ on edge $(ij)$ and similar for the mirror rapidities $v_k^{ij}$.
%and
%\beq d\mu(\textbf{v})e^{-\tilde{E}(\textbf{v}_{ij})\ell_{ij}} = \prod_{(ij)} n_{ij}! \prod_k dv^{ij}_k \mu({v^{ij}_k})e^{-\tilde{E}({v}^k_{ij})\ell_{ij}}.
%\eeq
Above $\mu(v)$ and  $\tilde{E}$ are respectively the phase space measure and the energy of the mirror particle and 
$\mathcal{H}^2(\textbf{v},\textbf{a})$ are the glued hexagon form factors with mirror particles labeled by $(\textbf{v},\textbf{a})$ inserted along the seams, the dependence on the external operators being left implicit. Below we provide explicit expressions for the case of interest, see \cite{wrappingthreeBGKV,wrappingshota} for general expressions. 

At weak coupling multiparticle mirror states are suppressed, both the energies 
%\beq
%e^{-\tilde{E}_a(v)} =  \frac{1}{x_v^{[+a]}x_v^{[-a]}}
%\eeq
and the measure being of~$O(g^2)$. Mirror contributions are also suppressed for large geometries so that at NNLO only edges with bridge length $\ell_{ij}=1$ can support mirror excitations. Moreover, at this order only one edge can be excited at a time: we may have an excitation in the adjacent or in the opposite edge to the non BPS operator.

Adjacent virtual corrections in the SU(2) sector provide an example of the selection rules restoration aforementioned. We now delve into this in detail to understand what is expected from the SOV formulas at NNLO. Structure constants in the SU(2) sector for states with R-charge $M$ when an adjacent bridge length $\ell<M$ vanish. At the classical level ($g=0$) this is simply the statement that we cannot contract $J$ scalars through a bridge of length $\ell$ \cite{tunneling,morphism,hexagons}. The asymptotic contribution $n_{ij} = 0$ correctly reproduce this selection rule at LO and NLO, but at NNLO a non-zero contribution is obtained when $\ell_{12}=1$. The claim is that the mirror factor cancels this contribution and restore the  symmetry \cite{spintwistsu2}. This reads
\beq
 \mathcal{A}_\ell^\texttt{asymptotic}  = - \mathcal{A}(n_{12}=1), \label{selectionrule}
\eeq
with
%\mathcal{A}(n_{ij}=0) =
%\beq
%\mathcal{A}(n_{ij}=0)=\sum_{\alpha\cup\bar{\alpha}} (-1)^{|\bar{\alpha}|}\frac{e^{-i p_{\bar{\alpha}}}}{h_{SU(2)}(\bar{\alpha},\alpha)} \label{asymptotic}
%\eeq
%and
%the asymptotic term $\mathcal{A}(n_{ij}=0)$ was defined in (\ref{carlosappendixformula}) and the single particle adjacent mirror contributions is given explicitly by 
\begin{align}
\mathcal{A}(n_{12}=1) = & \mathbb{N}_{J,\ell}  \int \frac{dv}{2\pi} \frac{a g^4}{\left(v^2 + a^2/4\right)^3} \vec{\tau}_{SU(2)}(v^\gamma,\mathcal{O}_1) \nonumber \times
\\ &\sum_{\alpha\cup\bar{\alpha}} (-1)^{|\bar{\alpha}|}\frac{e^{-i p_{\bar{\alpha}}} h_a(v^\gamma,\alpha)}{h^{SU(2)}_{\bar{\alpha},\alpha} h_a(\bar{\alpha},v^\gamma)}. \label{adjacent}
\end{align}
where $h_a$ are the fused hexagon dynamical factors, $v^\gamma$ denotes analytic continuation to mirror kinematics across the $x^{[+a]}_v$ cuts, $h^{SU(2)}_{\bar{\alpha},\alpha}$ are the SU(2) dynamical factors~(\ref{dynamicalfactor}) and
$ \vec{\tau}_{SU(2)}(v^\gamma,\mathcal{O}_1)$ is the forward SU(2) transfer-matrix eigenvalue for the Bethe state describing operator $\mathcal{O}_1$. We sum over partitions of the bethe state $\textbf{u} = \alpha \cup \bar{\alpha}$, and use the short hand notation $h_a(\bar{\alpha},v^\gamma)= \prod_{u_i \in \bar{\alpha}} h_a(u_i,v^\gamma)$
%\beq
%\nonumber
%h_a(\bar{\alpha},v^\gamma)= \prod_{u_i \in \bar{\alpha}} h_a(u_i,v^\gamma)%e^{-i p(u_i)} \nonumber
%\eeq
and similar for $h_{a}(v^\gamma,\alpha)$. To leading order the objects  in (\ref{adjacent}) read
\begin{align*}
h_a(v^\gamma,u_j)&=\frac{u_j+i/2}{ u_j+i/2-v+a i/2}\\
h_a(u_j,v^\gamma)&= \frac{-g_j^2}{u_j+i/2}\frac{(v-u_j)^2 + (a-1)^2/4}{(v-u_j) + i(a+1)/2}
\end{align*}
and $ \vec{\tau}_{SU(2)}(v^\gamma,\mathcal{O}_1)$ is given by 
\begin{align*}
 %\vec{\tau}_{SU(2)}(v^\gamma,\mathcal{O}_1)&=
 (a+1)\prod_{j=1}^J f_j^- + a  \prod_{j=1}^J f_j^+ + a \prod_{j=1}^J g_j f_j^-   + (a-1) \prod_{j=1}^J g_j f_j^+ \,
\end{align*}
where 
$g_j = (u_j-v-\tfrac{i}{2} (a+1))(u_j+\tfrac{i}{2})/((u_j-v-\tfrac{i}{2} (a-1))(u_j-\tfrac{i}{2}))$ and $f_j^\pm = (u_j-v + \tfrac{i}{2}(a\pm 1))/(u_j-\tfrac{i}{2} a)$.
%\beq \nonumber
%\!\!\text{where} \,\,\, f_j^\pm = \frac{u_j \pm i/2-v + i a/2}{u_j-i a/2}, \hspace{14pt} g_j = \frac{1-\frac{v+i a/2}{u_j-i/2}}{1-\frac{v+i a/2}{u_j+i/2}}.
%\eeq
%\begin{align*}
%&\vec{\tau}_{SU(2)}(v^\gamma,\mathcal{O}_1)= (a+1)\left(\prod_{j=1}^J \frac{x_v^{[-a]}-x_j^-}{\left(x_v^{[a]}\right)^{-1}-x_j^-}\right) + \\
%& a \left(\prod_{j=1}^J \frac{x_v^{[-a]}-x_j^+}{\left(x_v^{[a]}\right)^{-1}-x_j^-}\right) + a\left(\prod_{j=1}^J \frac{x_v^{[-a]}-x_j^-}{\left(x_v^{[a]}\right)^{-1}-x_j^-}\frac{1-x_v^{[+a]}/x_j^-}{1-x_v^{[+a]}/x_j^+}\right) +\\&(a-1)\left(\prod_{j=1}^J \frac{x_v^{[-a]}-x_j^+}{\left(x_v^{[a]}\right)^{-1}-x_j^-}\frac{1-x_v^{[+a]}/x_j^-}{1-x_v^{[+a]}/x_j^+}\right) \end{align*}
Equation (\ref{selectionrule}) holds off-shell. The LHS is a rational function of the rapidities $u_k$. The RHS integral can be evaluated by residues. Poles at $v=u_k \pm \tfrac{i}{2} (a-1)$ cancel after summing over the bound-state index while those at $v=\pm \tfrac{i}{2} a$ sum to a rational function matching the LHS.

Performing an R-symmetry transformation permuting the polarizations of operators $\mathcal{O}_2 \leftrightarrow \mathcal{O}_3$ should leave the structure constant invariant. In the integrability description this amounts to the replacement $\ell \leftrightarrow L-\ell$. The structure constant must therefore also vanish when $\ell = L-1$. The story in this case is more interesting. First, the LO and NLO asymptotic contributions only vanish on-shell, since after all the expression (\ref{mathcalA}) only knows that the right bridge $\ell_{31} = 1$ through the Bethe roots which solve Bethe equations on a chain of size $L = \ell_{12}+\ell_{31}$. At the NNLO the rational result is non-zero on-shell. The mirror contribution $\mathcal{A}(n_{31}=1)$ is now given by an expression identical to (\ref{adjacent}) with the last line replaced by 
\begin{align}
%\mathcal{A}(n_{31}=1) = &\mathbb{N}_{J,\ell} \int \frac{dv}{2\pi} \frac{a g^4}{\left(v^2 + a^2/4\right)^3} \vec{\tau}_{SU(2)}(v^\gamma,\mathcal{O}_1) \nonumber \times
%\\ &
\sum_{\alpha\cup\bar{\alpha}} (-1)^{|\bar{\alpha}|}\frac{e^{-i p_{\bar{\alpha}}} h_a(v^\gamma,\bar{\alpha})}{h^{SU(2)}_{\bar{\alpha},\alpha}h_a({\alpha},v^\gamma)}. \label{adjacentright}
\end{align}
After summing over the bound state index in the wrapping corrections we now obtain a complicated expression full of $\zeta$ and Polygamma functions whose coefficients vanish on-shell so that in the end the we are left with simple algebraic numbers that cancel the asymptotic result, reproducing the selection rule.
%In the next section we reproduce this on-shell vs off-shell selection rule structure from the SOV formulas through a dressing of the Q-functions by the geometry. Before shifting gears, lets consider the contribution from the virtual corrections around the seam opposite to the NBPS operator. As discussed are only present (at NNLO) when the operators $\mathcal{O}_2$, $\mathcal{O}_3$ are small enough so that $L_2 + L_3 - L_1 = 2$.
%Bottom discussion to add later.

{\begin{center}{\textit{Part II: dressing the Q-functions}} \end{center}}
\label{Part2}
Wrapping effects at NNLO are due to virtual particles propagating over bridges of size one. Since nontrivial SU(2) operators have R-charge $M\geq2$, NNLO wrapping effects are only present when restoring the selection rules discussed in Part I of \ref{Part1}.
There are two cases to consider in the SOV proposal: $\ell_{12} \equiv \ell=1$ and $\ell_{31} = 1$ i.e. $\ell=L-1$. The selection rules are realized trivially for $\ell=1$ simple due to the binomials in (\ref{pairingsu2}). This holds off-shell, as in the hexagon method. 
We henceforth focus on the nontrivial case. In other words, we seek to restore the important $\ell \leftrightarrow L-\ell$ symmetry for any possible lengths.

The NNLO SOV expression for structure constants in the SU(2) sector (\ref{su2structure}) was derived from the morphism action, (\ref{morphismaction}). Note that $\mathcal{A}_\theta$ depends only on $\theta_i$ with $i \leq \ell$. One might therefore naively think that the method is unaware of $\ell_{31}$ being short or long and therefore cannot distinguish when mirror corrections are relevant. 
The solution is provided by the \textit{right boundary} terms, i.e. terms with explicit $\partial_L$ and $\partial_{L-1}$ dependence in the second line of (\ref{morphismeq}) \cite{rightbound}. These can be ignored whenever $\ell<L-1$. Equation (\ref{morphismaction}) then reproduces the asymptotic structure constants. However, when $\ell = L-1$ the right boundary acts non-trivially. Its action generates the extra terms proportional to $\delta_{\ell,L-1}$ in (\ref{dressing}). Once these terms are properly taken into account, the selection rules are correctly reproduced when (\ref{su2structure}) is evaluated on-shell. Note that this corresponds to an infinite number of constrains on the on-shell action of the right boundary terms, and therefore the match is quite non-trivial.


\begin{thebibliography}{99}
\bibitem{bottom}
In this section we assume that $\ell_{23}\geq2$ so that bottom mirror corrections can be neglected \cite{wrappingthreeBGKV}. Otherwise, we expect its inclusion to follow a similar prescription as in the SL(2) case.
\bibitem{transcendentality}
It would be desirable to restore naive trascendentality counting, see also discussion at the end of appendix \ref{rootscharges}.
\bibitem{straight}
One simply use the Baxter equation (\ref{baxterT}) to solve for $\mathbb{Q}^{[+2]}$ in terms of $\mathbb{Q}$ and $\mathbb{Q}^{[-2]}$, plug the result in (\ref{simplifiedBaxter}), use recurrence identities to cancel the Harmonic functions and note that that what is left is zero provided $\mathbb{Q}$ is an even function. This is the case for twist 2 Bethe states.
\bibitem{techremark}
For $\ell_B>1$ the right hand side of (\ref{bernouli2}) is a rational number; the $\pi$'s and the $\zeta(3)$ in the second line are simply cancelling the $\pi$'s and the $\zeta(3)$ generated by the integral. For $\ell_B=1$ the $\pi$'s still cancel but there is no $\zeta(3)$ in the second line. (In that sense the expression is even simpler in this case when wrapping is present!) A $\zeta(3)$ term is generated by the integral so that the rhs of (\ref{bernouli2}) is of the form $\texttt{rational}+\texttt{rational}' \zeta(3)$ as it should.
\bibitem{spintwistsu2}
Note that primary SU(2) operators have $M\geq 2$ and therefore all structure constants for $\ell_{12} = 1$ vanish.
\bibitem{rightbound}
The action of $\partial_{L}$ boundary terms is irrelevant for the non-extremal correlators considered here.
\end{thebibliography}
\end{document}